\documentclass[traditabstract]{aa}
\usepackage{amssymb}
\usepackage{threeparttable}  
\usepackage{booktabs}
\usepackage{multirow}
\usepackage{bm}
\usepackage{amsfonts}
\usepackage{epstopdf}
\usepackage{longtable}
\usepackage{setspace} 
\usepackage{graphicx}
\usepackage{txfonts}
\usepackage{natbib}
\usepackage{hyperref}

\bibpunct{(}{)}{;}{a}{}{,} 
\begin{document}

\title{Calibrating the $\alpha$ parameter of convective efficiency using
    observed stellar properties}

\author{
    X.S. Wu\inst{1,2,3}     \and
    S. Alexeeva\inst{4}   \and
    L. Mashonkina\inst{4} \and
    L. Wang\inst{1,2}     \and
    G. Zhao\inst{1} \and
    F. Grupp\inst{2,5}    
}

\offprints{F. Grupp; \email{frank@grupp-astro.de}}
\institute{
    Key Laboratory of Optical Astronomy,
    National Astronomical Observatories,
    Chinese Academy of Sciences,
    20 Datun Road, Chaoyang District, Beijing 100012, China
    \and
    Max-Planck-Institut f\"ur Extraterrestrische Physik,
    Giessenbachstrasse, D-85748 Garching, Germany
    \and
    University of the Chinese Academy of Sciences,
    19A Yuquan Road, Shijingshan District, Beijing, 100049, China
    \and
    Institute of Astronomy, Russian Academy of Sciences,
    RU-119017 Moscow, Russia
    \and 
    Universit\"ats Sternwarte  M\"unchen,
    Scheinerstr. 1, D-81679 M\"unchen, Germany
}

\date{Received ; accepted}

 
\abstract
{
    Synthetic model atmosphere calculations are still the most commonly used tool
    when determining precise stellar parameters and stellar chemical
    compositions.
    Besides three-dimensional models that consistently solve for hydrodynamic
    processes, one-dimensional models that use an approximation for convective
    energy transport play the major role.
}
{
    We use modern Balmer-line formation theory as well as spectral energy
    distribution\,(SED) measurements for the Sun and Procyon to calibrate the
    model parameter $\alpha$ that describes the efficiency of convection in our
    1D models.
    Convection was calibrated over a significant range in parameter space,
    reaching from F-K along the main sequence and sampling the turnoff and giant
    branch over a wide range of metallicities.
    This calibration was compared to theoretical evaluations and allowed an
    accurate modeling of stellar atmospheres.
}
{
    We used Balmer-line fitting and SED fits to determine the convective
    efficiency parameter $\alpha$.
    Both methods are sensitive to the structure and temperature stratification
    of the deeper photosphere.}
{
    While SED fits do not allow a precise determination of the convective
    parameter for the Sun and Procyon, they both favor values significantly
    higher than 1.0.
    Balmer-line fitting, which we find to be more sensitive, suggests that the
    convective efficiency parameter $\alpha$ is $\approx 2.0$ for the main
    sequence and quickly decreases to $\approx 1.0$ for evolved stars.
    These results are highly consistent with predictions from 3D models.
    While the values on the main sequence fit predictions very well,
    measurements suggest that the decrease of convective efficiency as stars
    evolve to the giant branch is more dramatic than predicted by models.
}
{}


\titlerunning{CMA calibration}
\authorrunning{Wu et al.}

\maketitle

\section{Introduction}
    One-dimensional\,(1D) stellar model atmospheres are widely used for
    determining basic physical parameters, such as effective temperatures,
    surface gravities, luminosities, masses, and chemical compositions of stars
    at different evolutionary stages.
    They also serve as boundary conditions for stellar evolution codes.

    While codes for 3D model atmospheres self-consistently solve for convection
    in the atmospheric layers, 1D codes apply a simplified approach such as
    described by \citealt{Bohm1958} (BV hereafter) or \citealt{Canuto1991,
    Canuto1992} (CMA hereafter).
    As these formulations do not self-consistently determine the convective
    flux, at least one calibration parameter is required: the $\alpha$-parameter,
    which sets the efficiency of convection.

    Recent calibrations of the 1D $\alpha$-parameter based on 3D theoretical
    model atmospheres have been presented by \citealt{Magic2014} (hereafter,
    Magic14) and earlier by \cite{Ludwig1999}.
    The primary goal of this paper is to compare observational constraints on
    $\alpha_{\rm CMA}$, the $\alpha$-parameter in the formulation of
    \cite{Canuto1992}, to theoretical predictions of 3D models.
    This will allow us to use a proper $\alpha_{\rm CMA}$ for 1D model
    atmospheres across a wide range of stellar parameter space.

    We use the approach as \citealt{Fuhrmann1993} (FAG93 hereafter), using the
    dependence of Balmer line profiles on the atmospheric structure, which we
    call the $\alpha_{\rm CMA}$-parameter.
    For this test we use a sample of stars (including the Sun) with stellar
    parameter start values determined independently of model atmospheres, for
    example, by astrometrical methods.
    As another tool for determining $\alpha_{\rm CMA}$ , we use stellar spectral
    energy distributions (SED) in the visible and near-UV spectral range.
    This method is of course limited to objects with good absolute calibration
    of the SED.
    We here use the Sun and Procyon.

    On the model side, the adopted model atmosphere code is the plane-parallel,
    chemically homogeneous, local thermodynamical equilibrium\,(LTE) code
    MAFAGS-OS presented by \citet{Grupp2004a}.
    Stellar convection is treated in MAFAGS-OS according to the formalism of CMA.
    The authors take into account a spectrum of turbulent eddies with different
    sizes, which is more advanced  than the one single eddy in the mixing-length
    theory \citep[MLT;][]{Bohm1958}.
    The molecular weight within the convective zone is variable, a detail
    neglected by MLT.
    Nevertheless, it is a parametric and simplified approach, not a
    self-consistent numerical formalism.
    MLT furthermore assumes that the turbulence is incompressible and sets the
    mixing-length $\Lambda=\alpha H_{\rm p}$, where $H_{\rm p}$ denotes the local
    pressure scale height, and leaves the so-called mixing-length parameter
    $\alpha$ (hereafter, $\alpha_{\rm MLT}$) as a free parameter.
    Therefore, calibrations by comparing the models with the observational data
    are necessary.

    Some previous works suggest that $\alpha_{\rm MLT}$ is greater than unity
    for the Sun and varies with stellar parameters.
    For example, \citet{Miglio2005} constrained $\alpha_{\rm MLT}$ of the
    $\alpha$ Centauri A+B systems using a method from asteroseismology and found
    an $\alpha_{\rm MLT}$ for component B higher by 10\% than that of A.
    Moreover, $\alpha_{\rm MLT}$ was found to be strongly dependent on stellar
    mass \citep[e.g.,][]{Yildiz2006}, but not on metallicity
    \cite[e.g.,][]{Ferraro2006}.
    In contrast, using the solar-like stars in the Kepler field,
    \citet{Bonaca2012} indicated that $\alpha_{\rm MLT}$ correlates more
    significantly with metallicity than $T_{\rm eff}$ or $\log{g}$.
    With 2D hydrodynamical models of dwarf stars, \citet{Ludwig1999} showed that
    $\alpha_{\rm MLT}$ decreases with increasing $T_{\rm eff}$ and $\log{g}$, and
    this was confirmed by 3D simulations of \cite{Trampedach2007}.

    \citet{Canuto1992} introduced $\Lambda=\alpha z$ by taking into account local
    effects, where $z$ is the distance to the top of the convective zone, and its
    value is close to the polytrope $H_{\rm p}$.
    $\alpha=1$ correspond to the nonlocal $\Lambda=z$ in \citet{Canuto1991}.
    In previous versions of the MAFAGS-OS program, the $\alpha$ according to the
    CMA theory (hereafter, $\alpha_{\rm CMA}$) is fixed to 0.82, because the
    evolutionary stage of the Sun, as determined by its internal structure, can
    be well reproduced by this value \citep{Bernkopf1998}.
    However, this is only one calibration point in parameter space, there is no
    evidence that stars throughout the whole H-R diagram necessarily have the
    same $\alpha_{\rm CMA}$ values.

    In this paper, we analyze a sample of extensively studied "standard stars"
    \footnote{FG would like to suggest defining a "standard star" as a star that
    - as soon as it is studied in detail - is found to be pretty "non-standard".
    Procyon and HD\,140283 are stars that fit this category very well.}
    with different chemical compositions and evolutionary stages.
    All of them have angular diameter data obtained with modern interferometers,
    which -- together with distance measurements from \textsc{Hipparcos} --
    enable direct measurements of stellar radii, effective temperature, and
    surface gravities.
    The variation of $\alpha_{\rm CMA}$ across the H-R diagram can therefore be
    constrained.

\section{Stellar samples and basic stellar parameters}
\begin{figure}[ht]
            \centering
            \includegraphics[width=9cm]{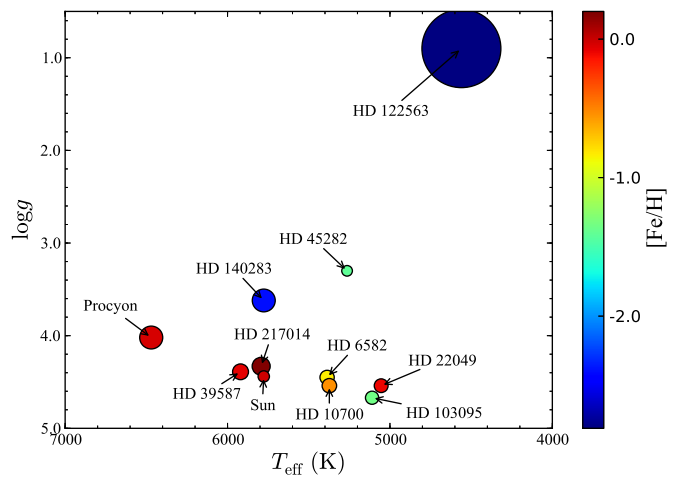}
            \caption{HRD of the program stars. Color encodes metallicity, size
                encodes stellar diameter measurements.}
           \label{hrd}
    \end{figure}

    Our sample consists of ten dwarf and turnoff stars as well as one giant, with
    effective temperatures ranging from 4600 K to 6600 K and metallicities
    ([Fe/H]) from $-$2.60 to +0.60.
    All objects have been observed with the Fiber Optics Cassegrain Echelle
    Spectrograph \citep[FOCES;][]{Pfeiffer1998} on the 2.2m telescope at Calar
    Alto Observatory.
    The optical spectra have signal-to-noise ratios of $>$200 and resolving
    powers ($R=\lambda/\Delta\lambda$) of $\approx$60,000.
    Data reduction was performed using the FOCES EDRS software package
    \citep[see][]{Pfeiffer1998}.
    Normalization was performed manually.
    The sample is shown in the physical Hertzsprung-Russell-Diagram\,(HRD) in
    Fig.\,\ref{hrd}.
    Interferometric diameter determination for eight of the program stars is
    available (see Table\,\ref{table-params}).

    To minimize the different instrumental response effects, the spectroscopic
    parameters ($T_{\rm eff}$, $\log{g}$, [Fe/H] and $\xi$) were taken from
    \cite{Fuhrmann1998,Fuhrmann2004}, who derived them with the same FOCES
    spectra as we used in this study.
    The available interferometric observations provide accurate bolometric
    fluxes (${\cal F}_{\rm BOL}$) with typical uncertainties of $\sim$2\%
    \citep[e.g.,][]{Boyajian2012a}.
    Combined with the interferometrically measured stellar angular diameter
    ($\theta$) measurements with typical relative uncertainties of 1$\sim$3\%
    for our samples, the uncertainties of $T_{\rm eff}=2341({\cal F}_{\rm BOL}/
    \theta^2)^{1/4}$ are lower than 1.5\%, which is precise enough as
    cross-validations of the spectroscopic data.

    Therefore, whenever possible, we used interferometric $T_{\rm eff}$ and
    $\theta$ given by recent measurements with VLTI or CHARA to compose
    interferometric parameters.
    The spectroscopic and interferometric parameters are summarized in Table
    \ref{table-params}.

    The $\log{g}$ of stars that have interferometric measurements parameters was
    derived using the direct angular diameter of the star, its distance from the
    Hipparcos parallax \citep{vanLeeuwen2007}, and its mass, either from the
    astrometric measurements of a binary companion (for Procyon\,A), or by
    fitting the stellar evolutionary track in the H-R diagram.
    The parameters metallicity ([Fe/H]) and micro-turbulence velocity ($\xi$),
    which cannot be obtained by interferometric means, were kept the same as
    those in the spectroscopic group.
    The solar abundance mixture was taken from \cite{Lodders2009}.
    For the five low-metallicity stars in our sample with [Fe/H] < $-$0.60
    (HD\,6582, [Fe/H] = $-$0.83; HD\,103095, [Fe/H] = $-$1.35; HD\,122563,
    [Fe/H] = $-$2.60; HD\,45282, [Fe/H] = $-$1.50; and HD\,140283, [Fe/H] =
    $-$2.38), $\alpha$-elements enhancements of [$\alpha$/Fe] = +0.4 were
    adopted.

    A model of Vega was calculated to consistently derive the theoretical
    zero-point of the spectrophotometric data.

\begin{table*}
 \begin{minipage}{120mm}
 \renewcommand{\arraystretch}{1.5}
 \tiny{
\caption{Best-fitting $\alpha_{\rm CMA}$}
 \label{table-params}
 \begin{tabular}{ccccccccccccc}
    \hline
    \multirow{2}{*}{HD}                   &
    \multirow{2}{*}{$T_{\rm eff}$\,(K)}   &
    \multirow{2}{*}{$\theta_{\rm LD}$}    &
    \multirow{2}{*}{Ref.}                 &
    \multirow{2}{*}{$\log{g}$ $^a$}       &
    \multirow{2}{*}{[Fe/H]$^a$}           &
    \multirow{2}{*}{$T_{\rm eff}^*$\,(K)} & 
       \multicolumn{3}{c}{The best $\alpha_{\rm CM}$} &
       \multicolumn{3}{c}{$\chi^2_{\rm red}$(d.o.f)}  \\
       &           &  & & & & & H$_\alpha$ & H$_\beta$  & H$_\gamma$ & H$_\alpha$ & H$_\beta$  & H$_\gamma$ \\
    \hline
    \multicolumn{13}{c}{Interferometric and IRFM} \\
    \hline

    Sun    &5780$\pm40$  & --              &                       & 4.44 &   0.00 & 5780 & 2.0 & $\ge$1.0 &$\ge$1.0   &6.05 (244)&  --   &   -- \\
    Procyon&6530$\pm50$  & 5.448$\pm$0.053 & \cite{Kervella2004}   & 3.98 &$-$0.05 & 6524 & 1.9 & $\ge$1.0 &$\ge$1.0   &0.94 (309)&  --   &   -- \\
    10700  &5380$\pm50$  & 2.078$\pm$0.031 & \cite{DiFolco2004}    & 4.53 &$-$0.49 & 5330 & 1.8 & $\ge$1.0 &$\ge$1.0   &0.72 (195)&  --   &   -- \\
    103095 &4820$\pm100$ & 0.679$\pm$0.015 & \cite{Creevey2012}    & 4.59 &$-$1.30 & 5100 & 2.4 & 1.9      &$\ge$0.5   &4.22 (648)&1.84 (217)&   \\
    39587  &5960$\pm50$  & 1.051$\pm$0.009 & \cite{Boyajian2012a}  & 4.48 &$-$0.07 & 5960 & 1.9 & $\ge$1.0 &$\ge$0.5   &1.11 (187)&   --  &   -- \\
    6582   &5260$\pm85$  & 0.972$\pm$0.009 & \cite{Boyajian2012a}  & 4.51 &$-$0.83 & 5370 & 1.7 & $\ge$0.5 & --        &0.58 (112)&   --  &   -- \\
    217014 &5800$\pm55$  & 0.748$\pm$0.027 & \cite{Baines2008}     & 4.28 &  +0.20 & 5780 & 2.0 & $\ge$1.0 &$\ge$1.0   &1.08 (236)&   --  &   -- \\
    22049  &5120$\pm75$  & 2.148$\pm$0.029 & \cite{DiFolco2004}    & 4.61 &$-$0.09 & 5045 & 2.2 & $\ge$1.0 &--         &0.83 (252)&   --  &   -- \\
    140283 &5780$\pm50$  & --              & \cite{Casagrande2010} & 3.70 &$-$2.38 & 5780 & 1.7 & 1.0& 1.0 & 1.00 (201)&0.60 (547)&1.13 (264)    \\
    122563 &4600$\pm50$  & 0.948$\pm$0.012 & \cite{Creevey2012}    & 1.60 &$-$2.60 & 4600 & 1.0 & 0.5& 0.5 & 0.46 (533)&2.50 (236)&2.43 (80)     \\
    45282  &5300$\pm50$  & --              & \cite{Casagrande2010} & 3.07 &$-$1.50 & 5310 & 1.0 & 0.5&  -- & 0.96 (574)&0.71 (124)&     --       \\
    \hline  \multicolumn{13}{c}{Spectroscopic} \\ \hline
    Procyon & 6530 & -- & This work           & 3.96 & $-$0.05 & 6530& 1.9& $\ge$1.0 &$\ge$1.0 &   --     &0.99 (309) &  -- \\ 
    103095  & 5110 & -- & \cite{Fuhrmann1998} & 4.66 & $-$1.35 & 5110& 2.4& 1.9      &$\ge$0.5 &   --     &1.05 (648) & 1.07 (217)\\
    39587   & 5920 & -- & \cite{Fuhrmann2004} & 4.39 & $-$0.07 & 5920& 1.5& $\ge$0.5 &$\ge$0.5 &   --     &0.96 (187) &  -- \\
    6582    & 5390 & -- & \cite{Fuhrmann2004} & 4.45 & $-$0.83 & 5390& 2.3& $\ge$1.0 & --      &0.72 (112)&    --     &  -- \\
    217014  & 5790 & -- & \cite{Fuhrmann1998} & 4.33 &   +0.20 & 5790& 2.0& $\ge$1.0 &$\ge$1.0 &0.93 (236)&    --     &  -- \\
    22049   & 5050 & -- & \cite{Fuhrmann2004} & 4.54 & $-$0.09 & 5050& 2.2& $\ge$1.0 & --      &0.80 (252)&    --     &  -- \\ \hline 
 \end{tabular}  
 \begin{tablenotes}  
     \item[a] {\bf Notes.} $^a$ References of the adopted $\log{g}$ and [Fe/H]
         are in the text.
         $T^*_{\rm eff}$ denotes the best effective temperatures obtained from
         the fitting procedure.
 \end{tablenotes}
 }        
 \end{minipage}
\end{table*}       

\section{Model atmospheres}
    For each set of stellar parameters, model atmospheres were calculated using
    the model atmosphere code MAFAGS-OS \citep{Grupp2004a,Grupp2004b,Grupp2009}.
    The opacity-sampling (OS) code is based on the opacity distribution function
    (ODF) version of T.\,\cite{gehren79}. In the reprogrammed version of
    \citet{reile87}, the code relies on the following basic assumptions:
   \begin{itemize}
      \item Plane-parallel 1D geometry.
      \item Chemical homogeneity throughout the atmosphere.
      \item Hydrostatic equilibrium.
      \item Convection is treated according to the formalism of
          \citet{Canuto1991, Canuto1992}.
      \item No chromosphere or corona.
      \item Local thermodynamical equilibrium.
      \item Flux conservation throughout all 80 layers.
       \end{itemize}
    While these assumptions can break down for hot stars, stars with very
    extended atmospheres, and the coolest stars, they are valid for temperatures
    from 4000 through 15000 K and for gravities from the main sequence down to
    $\log{g}\approx0$. 

    The model was iterated until the flux was constant ($<0.5\%$ level)
    throughout the atmosphere.

    \subsection{Convection and atmospheric structure}

    \begin{figure}[ht]
            \centering
            \includegraphics[width=6cm]{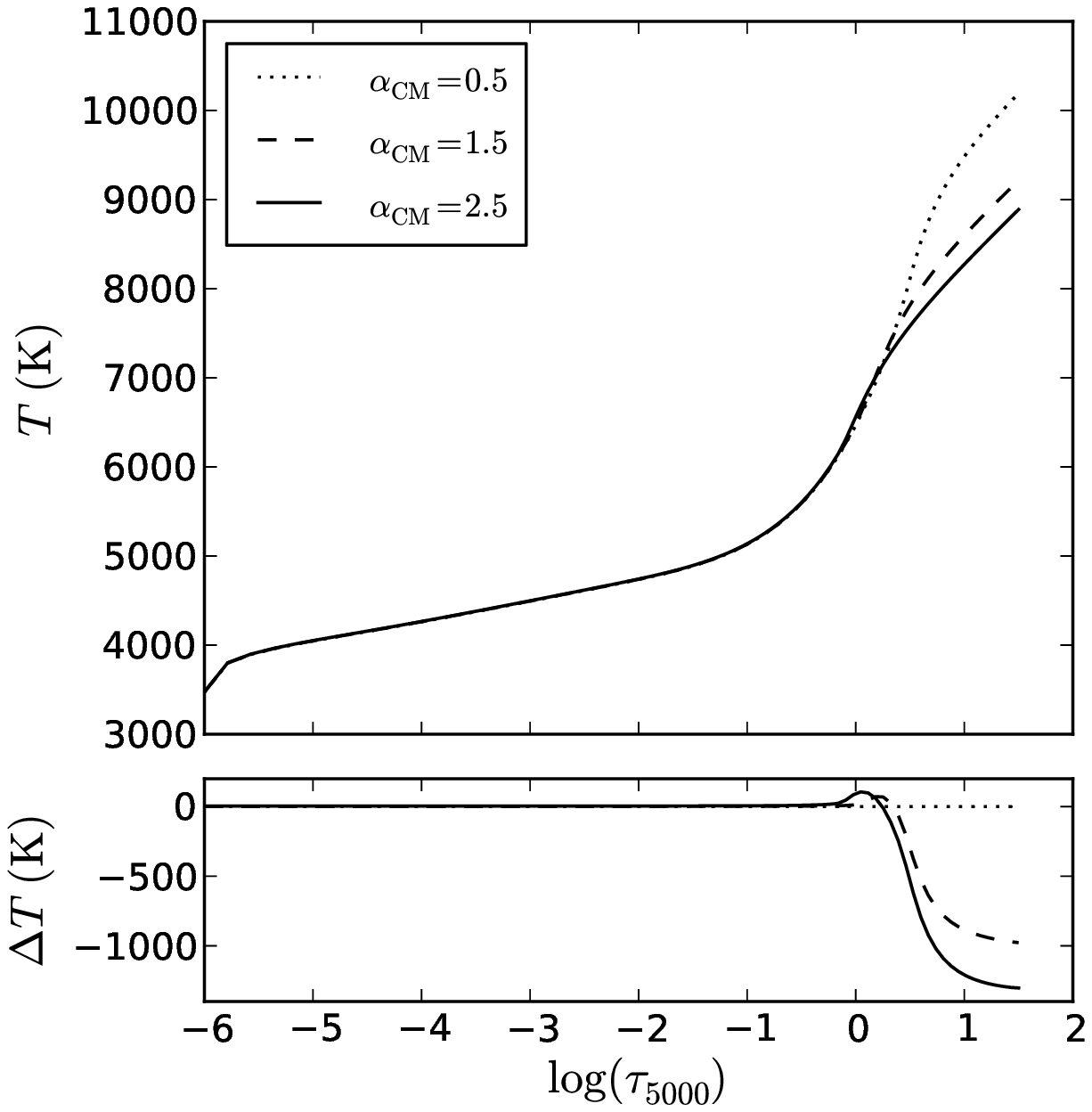}\\
            \includegraphics[width=6cm]{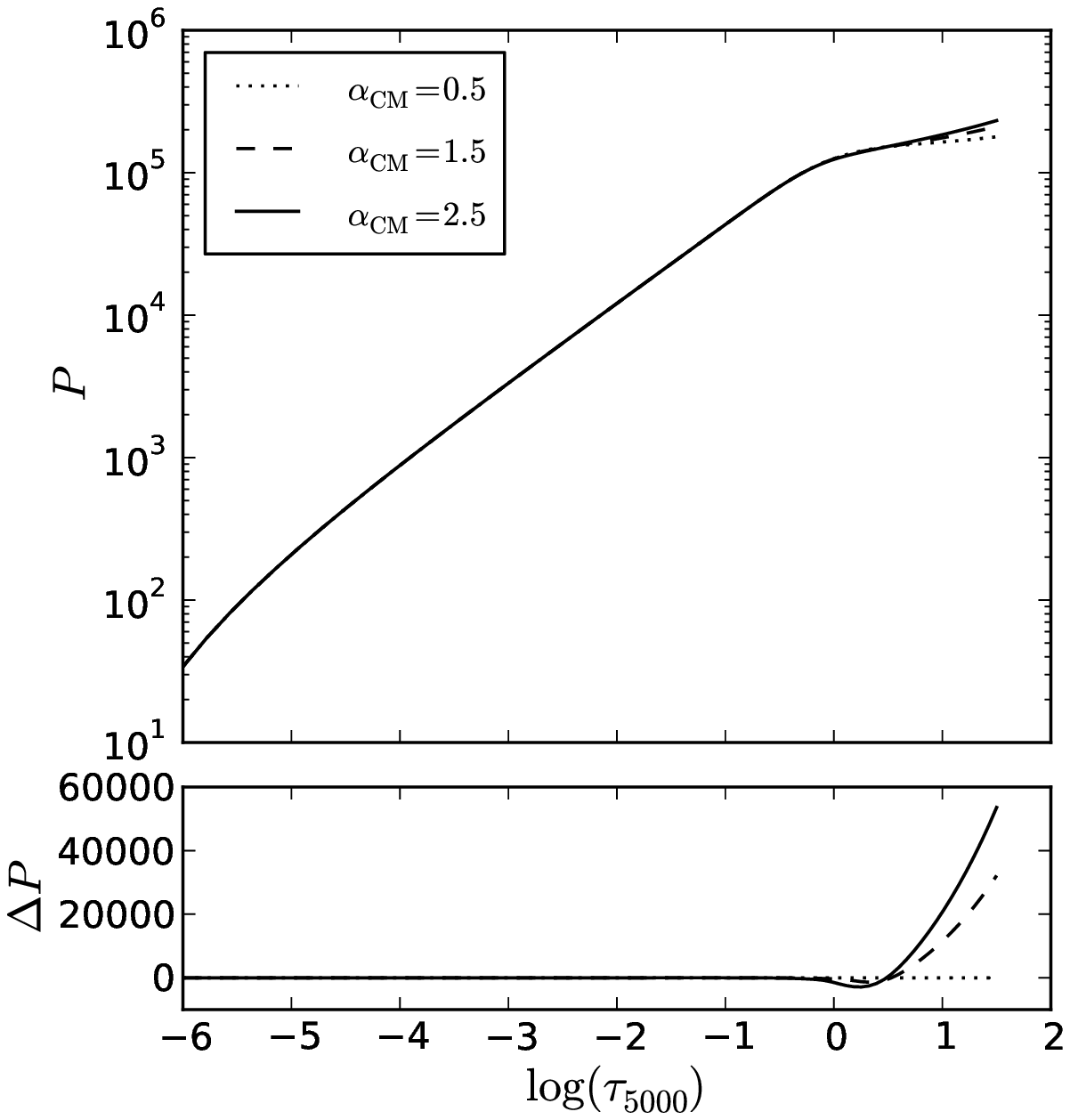}\\
            \includegraphics[width=6cm]{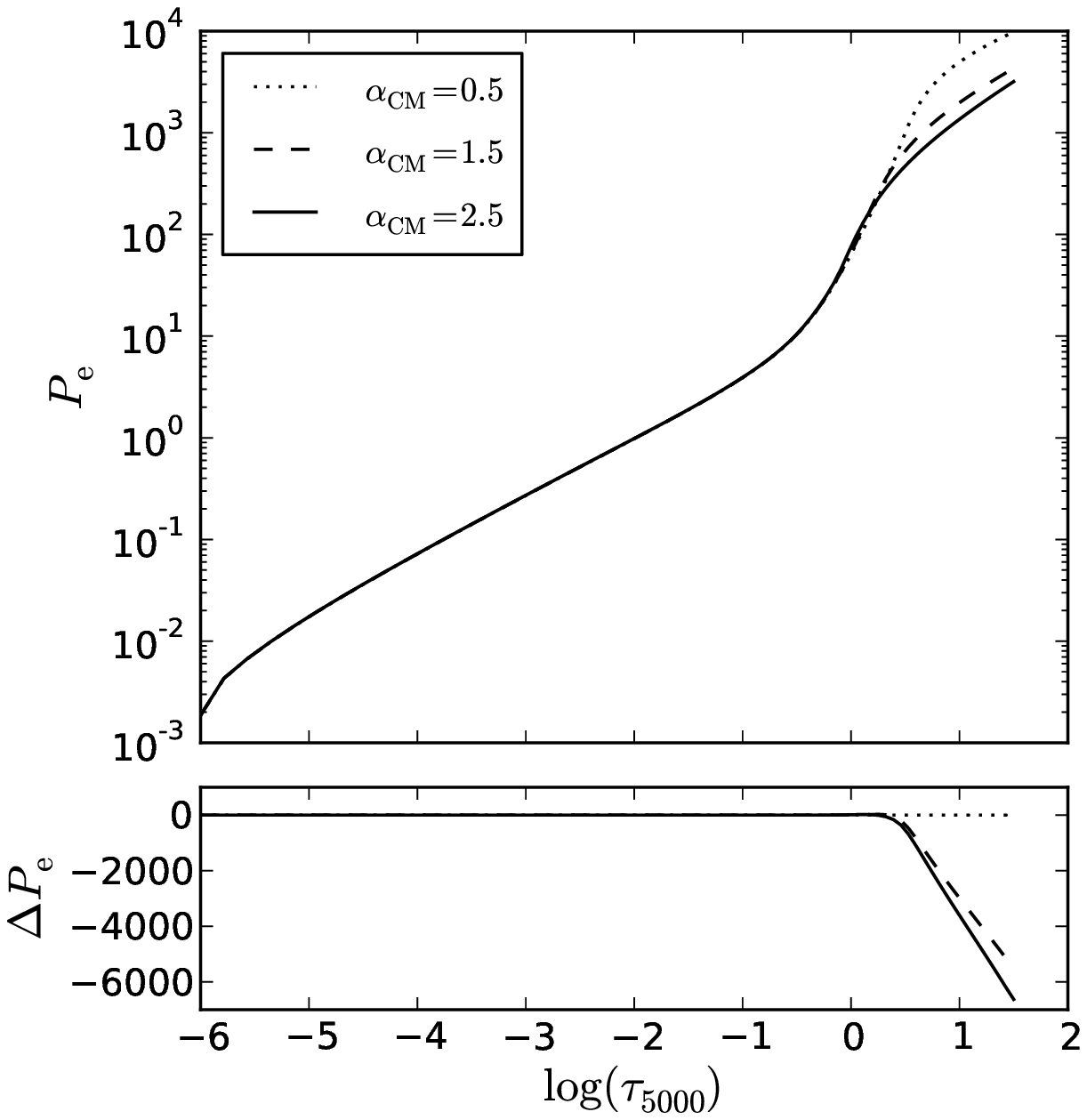}
            \caption{Temperature ($T$), pressure ($P$), and electron pressure
            ($P_{\rm e}$) versus optical depth $\log(\tau_{5000})$ with
            three different mixing-length parameters $\alpha_{\rm CMA}$ in
            the MAFAGS-OS models of Sun.}
           \label{fig_cma_structure}
    \end{figure}

    Figure\,\ref{fig_cma_structure} shows the influence of the convective
    efficiency parameter $\alpha_{\rm CMA}$ on temperature stratification, gas
    pressure, and electron pressure of a solar atmospheric model.
    While a low value of $\alpha_{\rm CMA}$ results in low convective flux, high
    $\alpha_{\rm CMA}$ leads to an increase of the convective flux and - as the
    total flux is preserved through all layers - lower radiative flux in the
    convection zone of the stellar atmosphere.
    Low radiative flux leads to lower temperature gradients.
    Therefore, the high-$\alpha_{\rm CMA}$ case shows a lower temperature in the
    inner region of the stellar atmosphere.

    For our program stars, increasing $\alpha_{\rm CMA}$ from 0.5 to 2.5 leads
    to decreasing temperatures in the inner atmosphere by up to 1000--1300 K at
    $\log(\tau_{5000})\simeq1.5$.

\begin{figure}[ht]
        \centering
        \includegraphics[width=9cm,clip=true,trim=6mm 0mm 1mm 0mm]{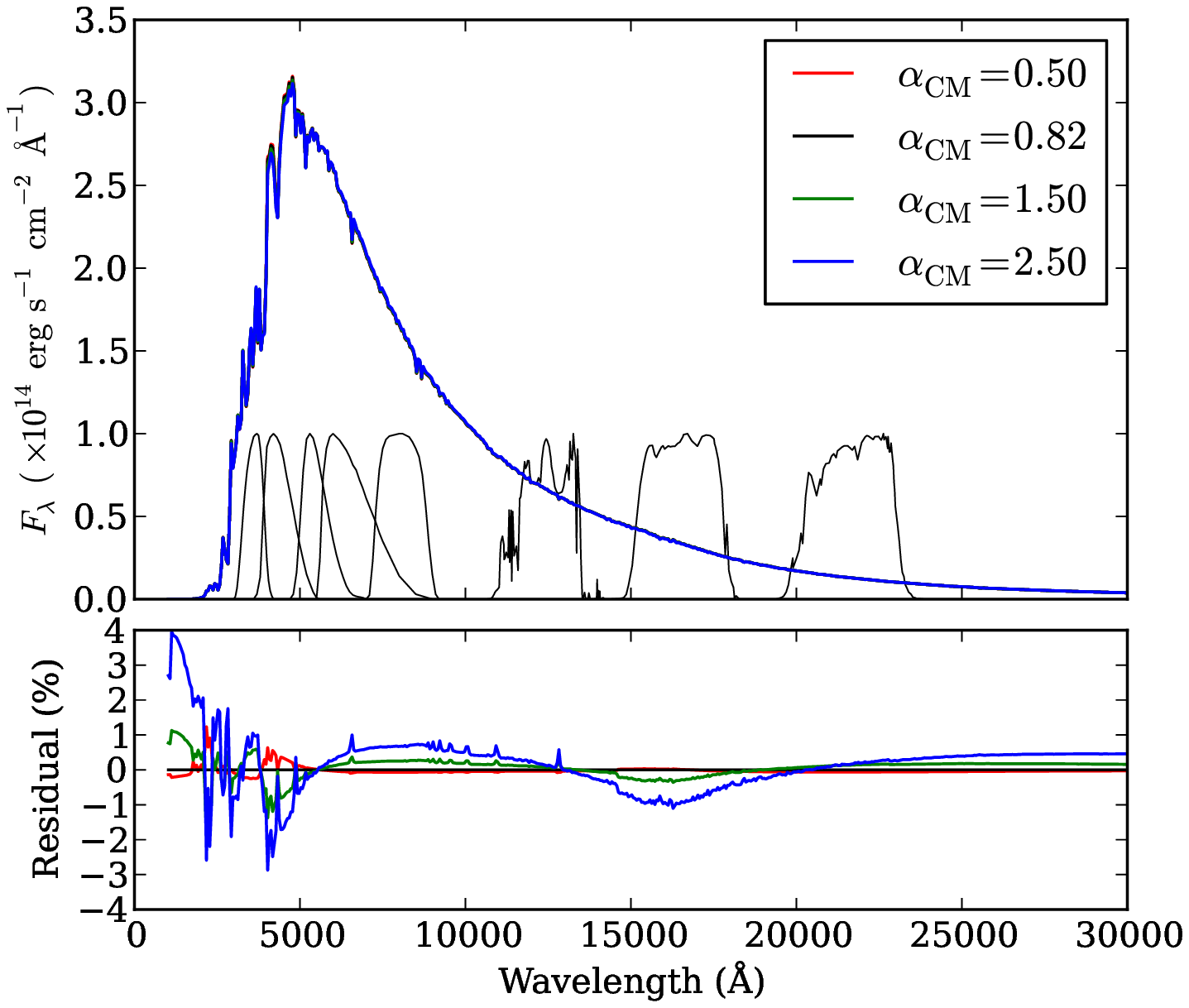}
        \caption{
            Upper panel: Computed solar flux $F_\lambda$ with different
            $\alpha_{\rm CMA}$.
            The spectrum is sampled every 50 \AA\ along the wavelength.
            The transmission curves of U, B, V, R, I in Bessell photometric
            system, and J, H, K$_{\rm s}$ in 2MASS system are also overplotted
            (from left to right).
            The residuals to fluxes with $\alpha_{\rm CMA}=0.82$ are plotted in
            the lower panel.}\label{fig_cma_flux}
    \end{figure}

    The effect of a variation of $\alpha_{\rm CMA}$ on the solar SED is shown in
    Fig.\,\ref{fig_cma_flux}.
    Flux changes of up to 5\% of the solar SED in the UV and of 1\% in the
    near-IR region are notable.
    There are some node points at 550, 1350, and 2050\,nm around which the
    fluxes are insensitive to mixing-length parameters, while the continuum of
    the SED between each two shows opposite tendencies with $\alpha_{\rm CMA}$.

    The trends in model structures and fluxes are similar for other stars in our
    sample, with the only exception of the A0V star Vega.
    As Vega has no convection zone reaching the stellar atmosphere, no change
    with $\alpha_{\rm CMA}$ is present in the stellar flux.

    \section{Spectral energy distributions}
    In the following section we directly compare the computed absolute fluxes to
    the observational data to calibrate $\alpha_{\rm CMA}$.
    Absolute flux data are only available for a few stars.
    For our sample, reliable absolute SED data are only available for the Sun
    and Procyon.

    \subsection{Near-UV -- visible -- near-IR domain}

    \begin{figure*}[ht]
        \centering
        \includegraphics[width=14cm]{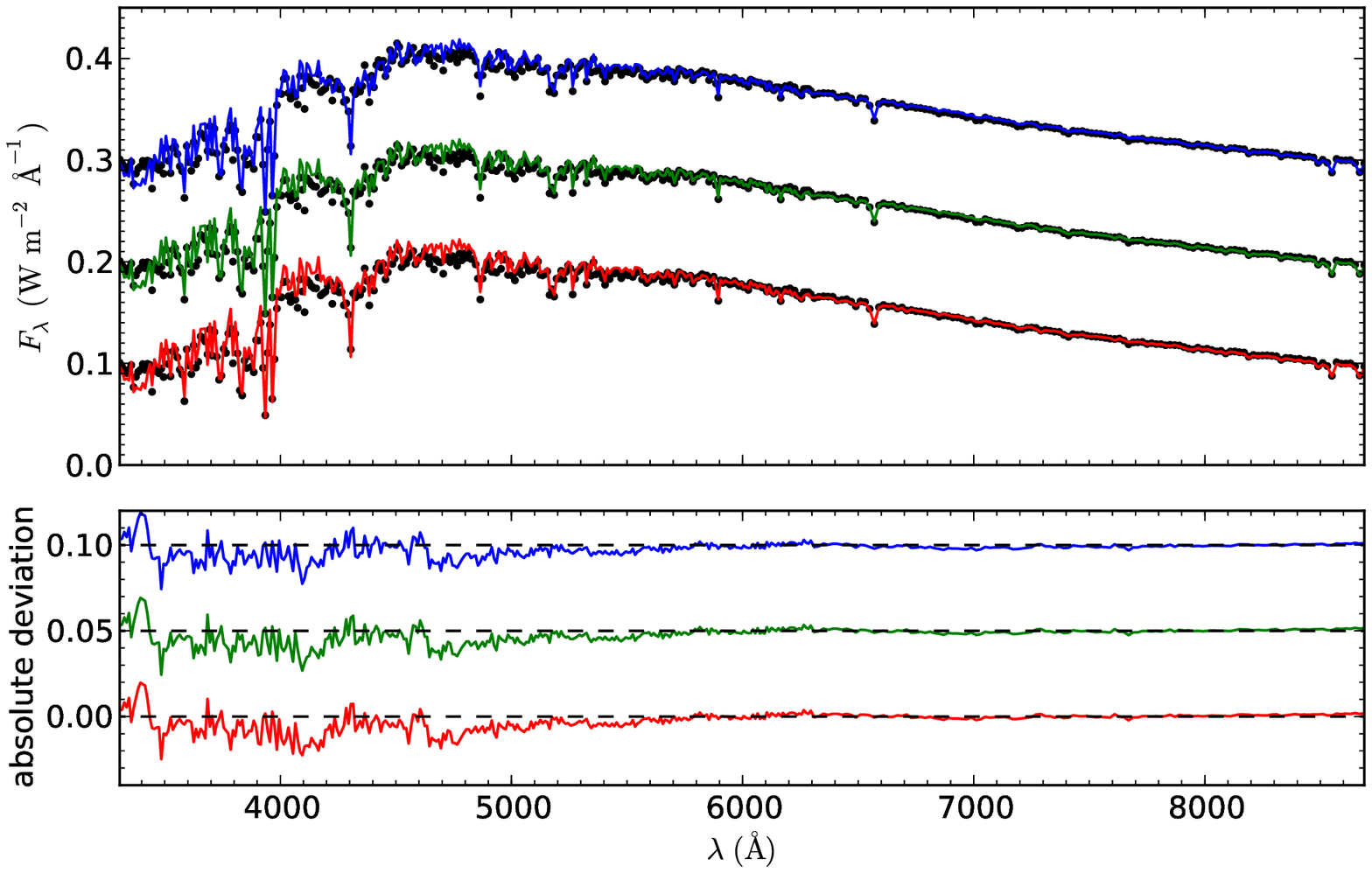}\\
        \includegraphics[width=14cm]{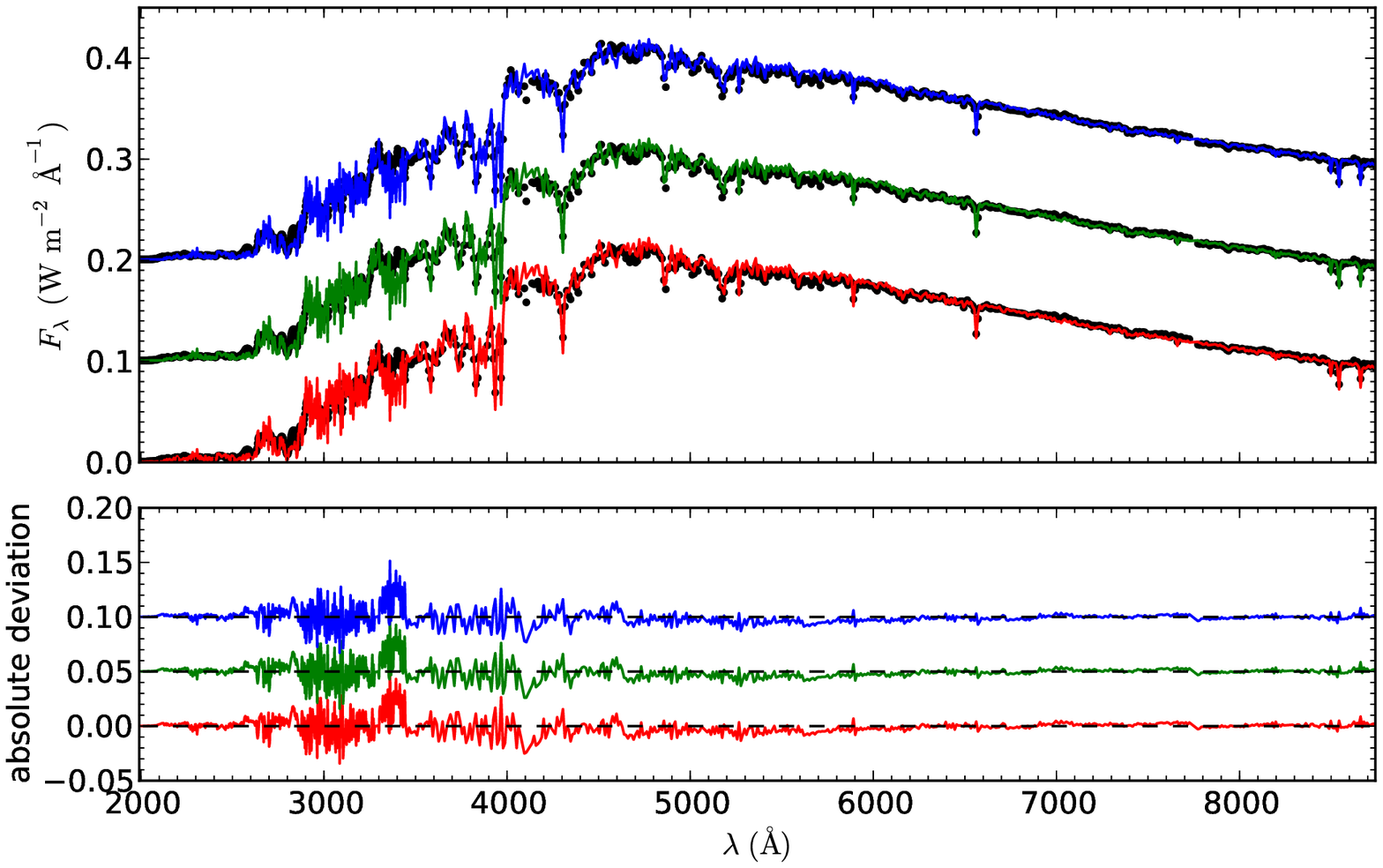}
        \caption{Comparison of solar absolute fluxes ({\it solid dots}) to
            synthetic flux levels with $\alpha_{\rm CMA}=0.5$ ({\it red}), 1.5
            ({\it green}), and 2.5 ({\it blue}).\newline
            Observational data were taken from \cite{Neckel1984} ({\it upper
            panel}) and \cite{Thuillier2003} ({\it lower panel}).
            The absolute deviation ($F_{\rm model}-F_{\rm obs}$) is plotted in
            the lower part of each panel.
            Offsets of 0.05 were added to improve the visualization.}
        \label{fig_sun_obs_visable}
    \end{figure*}

    \begin{figure*}[ht]
    \centering
        \includegraphics[width=6cm]{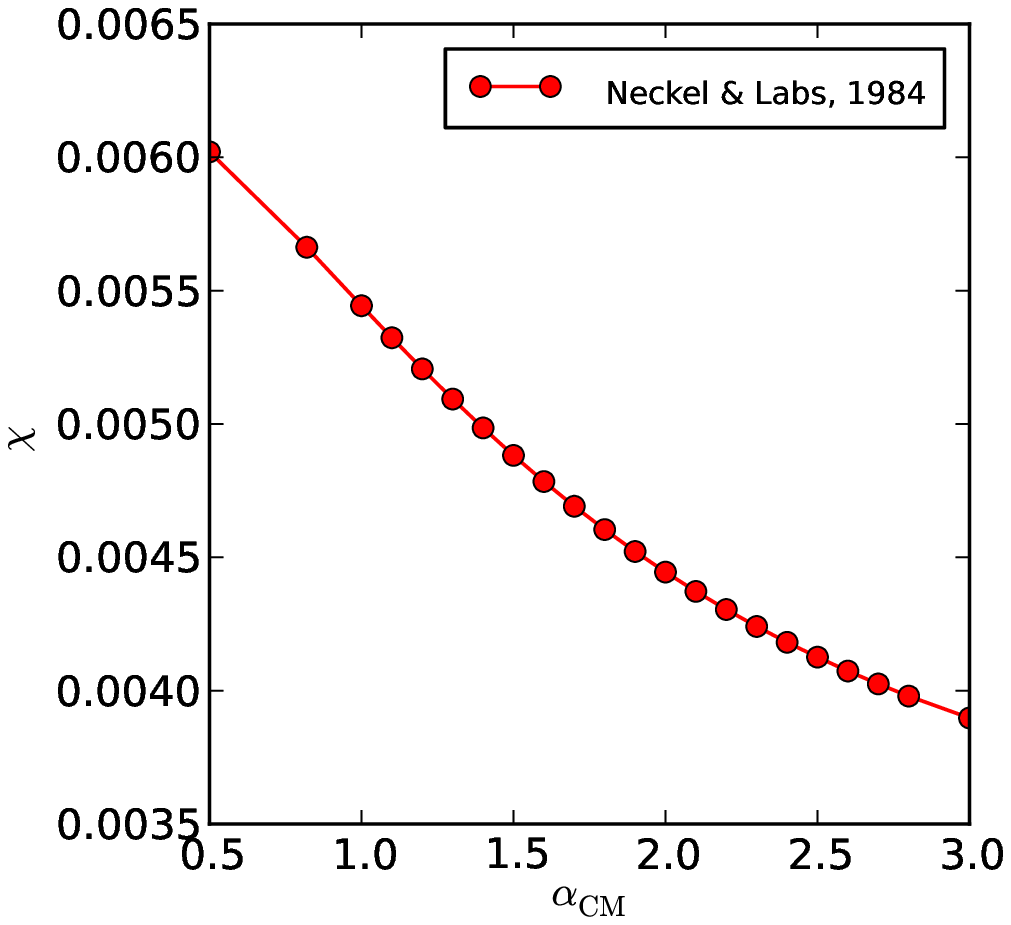}
        \includegraphics[width=6cm]{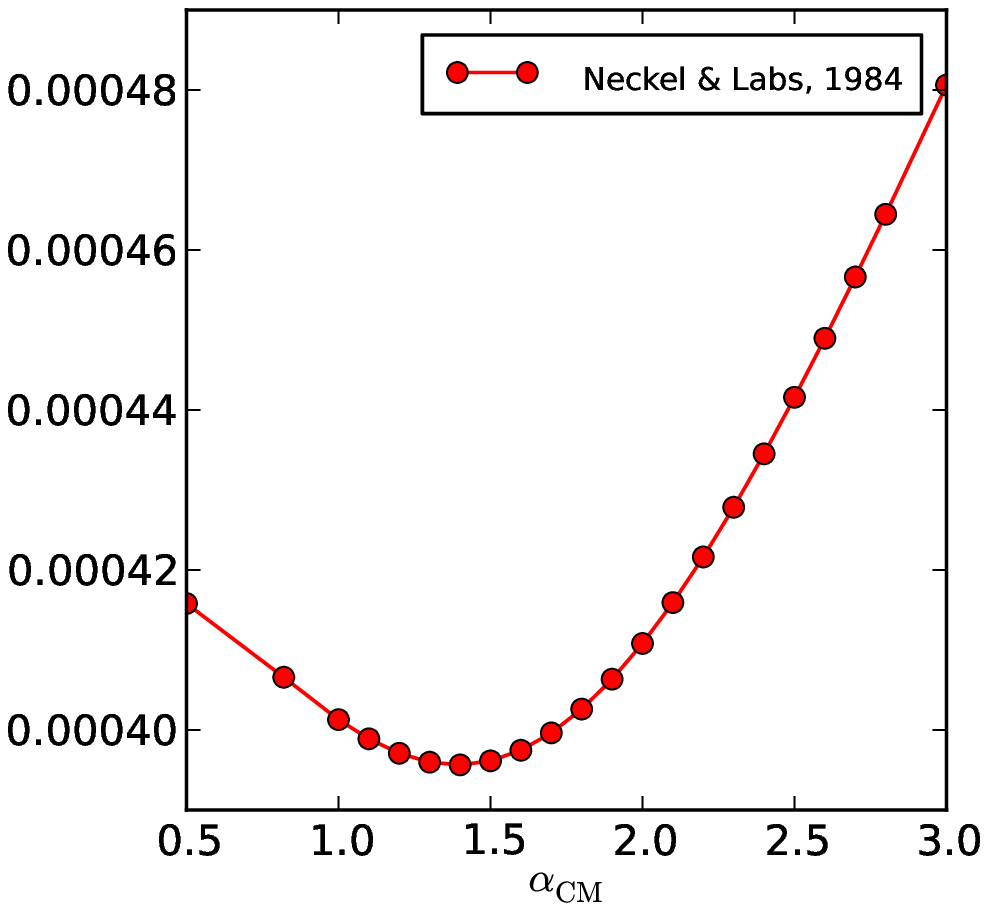}\\
        \includegraphics[width=6cm]{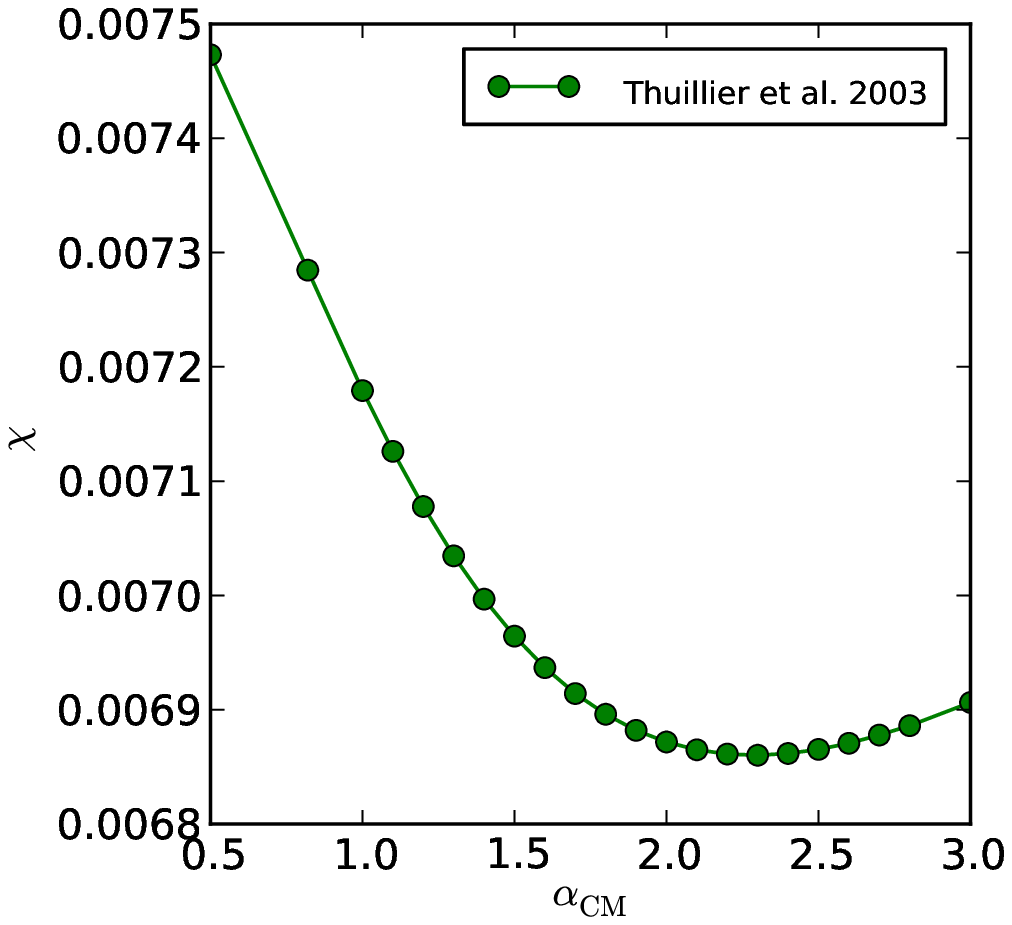}
        \includegraphics[width=6cm]{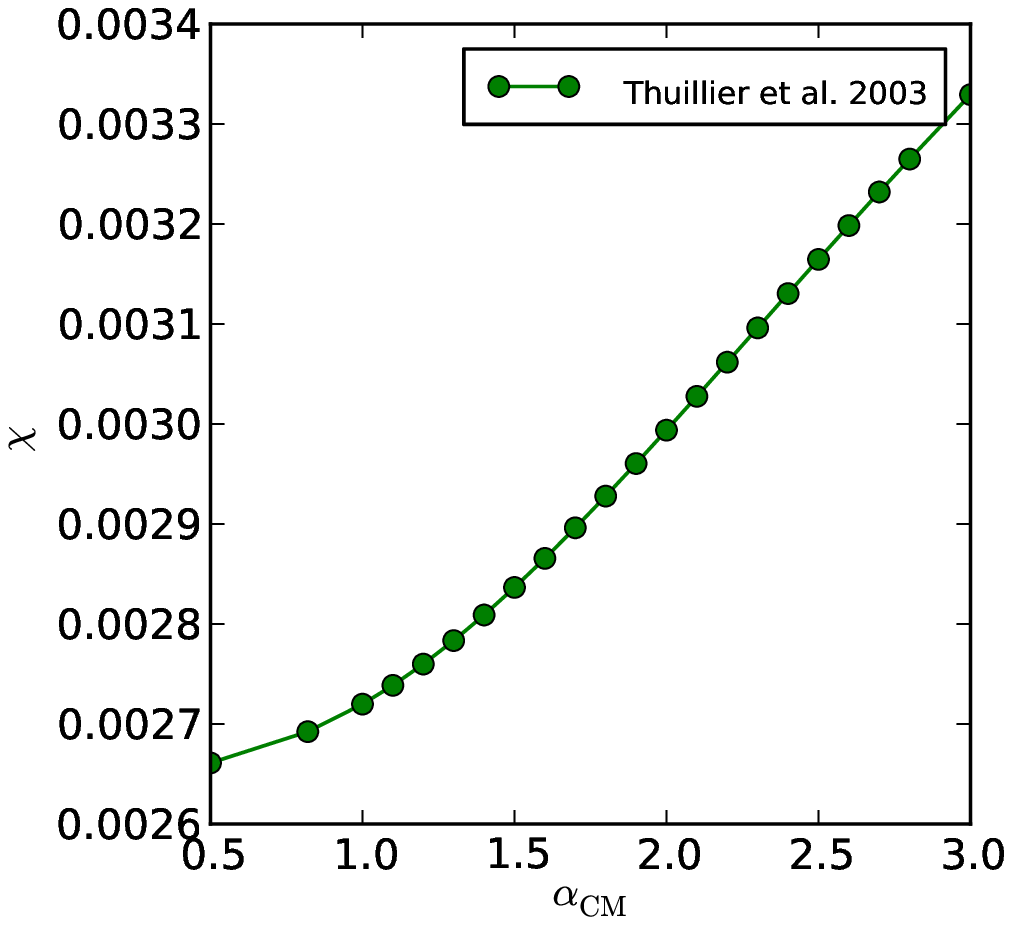}
        \caption{Relations of $\chi^2$ of observational and computed fluxes
            versus $\alpha_{\rm CMA}$ for the solar irradiance of
            \cite{Neckel1984} and \cite{Thuillier2003}.
            Observational data between 4500 and 8700 \AA\ (left plots) and
            between 5500 and 8700 \AA\ (right plots) were taken into account.}
        \label{fig_sun_obs_dev}
    \end{figure*}

    \cite{Neckel1984} measured solar irradiance between 3300 and 12500 \AA\ by
    combining the absolute fluxes of high-resolution spectra obtained with the
    Fourier Transform Spectrometer (FTS) at the McMath Solar Telescope on Kitt
    Peak.
    The spectra were averaged every 10\,\AA\ between 3305 and 6295\,\AA, and
    every 20\,\AA\ between 6310 and 8690\,\AA.

    \cite{Thuillier2003} presented the solar irradiance from 2000\,\AA\ to
    24000\,\AA\ obtained with the SOLSPEC and SOSP spectrometers mounted on
    space missions.
    However, for wavelength ranges above 8700\,\AA, their data are sampled with
    a rate different from that in visible and UV domains.
    Therefore we only adopted the spectra below 8700\,\AA\ to compare with our
    solar fluxes.

    Figure\,\ref{fig_sun_obs_visable} shows the data and absolute deviations of
    solar irradiance of \cite{Neckel1984} and \cite{Thuillier2003} together with
    the computed solar fluxes for three different $\alpha_{\rm CMA}$ values of
    0.5, 1.5, and 2.5.
    A visual check prefers a fit between blue ($\alpha_{\rm CMA}=2.5$) and green
    ($\alpha_{\rm CMA}=1.5$).

    The left and right panels of Fig.\,\ref{fig_sun_obs_dev} compare the measured
    and observed fluxes starting from 4500\,\AA\ and 5500\,\AA.
    The strong dependency on the blue end of the comparison window does not allow
    precisely determining $\alpha_{\rm CMA}$ from these data.
    Nevertheless, data points with values below 1.0 can be excluded from the
    Neckel data.
    The data of \citet{Thuillier2003} would only allow for very low convective
    efficiency if the blue region is ignored.
    In general, we find an overestimation of the solar flux in the regime below
    5000\,\AA.
    This can be caused by missing atomic or molecular opacity.
    Variations of convective efficiency do not change this region significantly,
    thus can be excluded as the cause of this shortcoming. 
     
    \subsection{Ultraviolet domain}

    \begin{figure}[h]
        \centering
        \includegraphics[width=9cm]{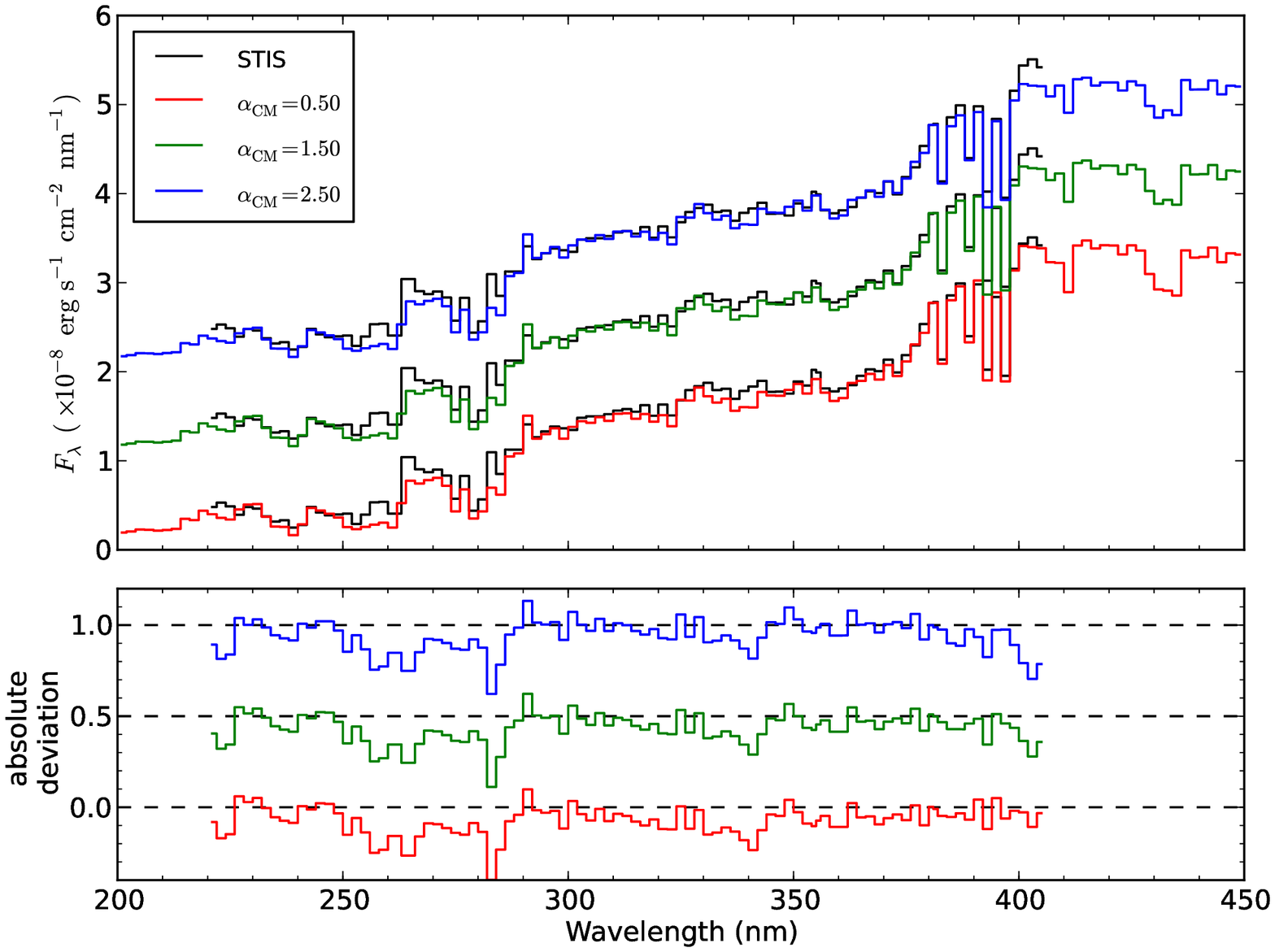}
        \caption{
            Comparison of absolute fluxes of Procyon\,A ({\it solid lines})
            obtained with HST/STIS with MAFAGS-OS model fluxes with
            $\alpha_{\rm CMA}=0.5$ ({\it red}), 1.5 ({\it green}), and 2.5
            ({\it blue}) in model (a).
            The fluxes were summed every 2 nm.
            The absolute deviation ($F_{\rm model}-F_{\rm obs}$) were plotted in
            the lower panel.
            Offsets of 1.0 in the upper panel and 0.5 in the lower panel were
            added for clarity.}
        \label{fig_stis_procyon}
    \end{figure}

    \begin{figure}[h]
        \centering
        \includegraphics[width=9cm]{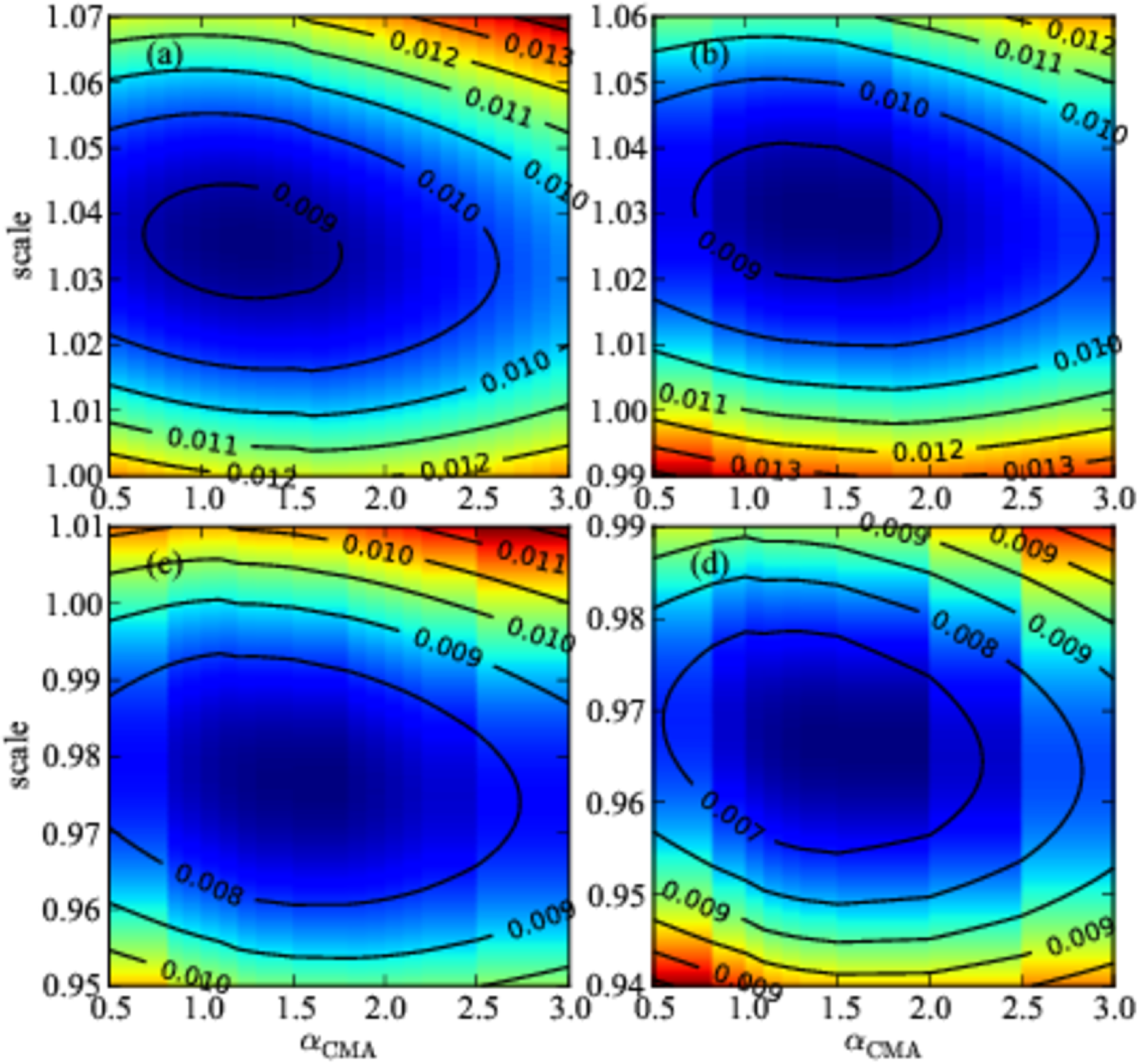}
        \caption{
            $\chi^2$ versus $\alpha_{\rm CMA}$ and the scaling factor for the
            HST/STIS data and the computed absolute fluxes between 220 and
            405\,nm with models of different $T_{\rm eff}$ for Procyon\,A.
            From (a) to (d): $T_{\rm eff}=6520/6530/6550/6590$ K.}
        \label{fig_stis_procyon_chi2}
    \end{figure}

    The ultraviolet spectrum of Procyon\,A has been obtained with the Space
    Telescope Imaging Spectrograph (STIS) onboard the Hubble Space Telescope
    (HST) with high-quality calibration in the range from 220\,nm up to 405\,nm.
    As the STIS spectra span several echelle orders and overlap between each
    pair of orders, we averaged bins of 2\,nm width for the STIS measured data
    and MAFAGS-OS model calculations to allow a most direct comparison.
    This comparison is shown in Fig. \ref{fig_stis_procyon} for $\alpha_{\rm CMA}$
    ranging from 0.5 to 2.5 for a model with a $T_{\rm eff}/\log{g}/$[Fe/H]$/\xi$
    = 6530/3.98/$-0.05$/2.10.
    MAFAGS-OS underestimates the fluxes by at most $\sim$35\% for \ion{Mg}{II}
    H \& K at 280\,nm.
    The average deviation is within 4\%.

    Figure \ref{fig_stis_procyon_chi2} displays the 2D plots of $\chi^2$ versus
    $\alpha_{\rm CMA}$ (x axes) and the scaling factor of computed flux (y-axes)
    for four models with different temperatures.
    The difference of $\sim70$ K in $T_{\rm eff}$ causes a difference of
    $\sim$4\% in the total fluxes.
    The scaling factors were introduced to account for the offsets between the
    computed fluxes and the observational data to first-order approximation.
    The relative uncertainty ($\sim1\%$) of the angular diameter can lead to an
    offset of up to 2\% of the observational SED.
    Therefore the $\pm3\%$ of the scaling factor on the total flux is reasonable.
    The derived $\alpha_{\rm CMA}$ for each model was estimated by minimizing
    the $\chi^2$ with the observational SED over the 2D plane.
    We found that model (c) and (d) led to smaller $\chi^2$ with the observational
    data, which favors higher $T_{\rm eff}$ for Procyon\,A.
    Models (b), (c), and (d) led to quite consistent $\alpha_{\rm CMA}\simeq1.4$
    to minimize $\chi^2$, while model (a) favors smaller $\alpha_{\rm CMA}\simeq
    1.3$.


\section{NLTE calculations of Balmer-line profiles}

    Our analysis is based on the nonlocal thermodynamical equilibrium (NLTE)
    line formation for \ion{H}{i} using the method described in
    \citet{Mashonkina2008}.
    In brief, the model atom includes levels with principal quantum numbers up
    to $n\le19$.
    We checked the influence of inelastic collisions with neutral hydrogen
    atoms, as computed following \citet{Steenbock1984}, on the statistical
    equilibrium (SE) of \ion{H}{i} and resulting profiles of the Balmer lines in
    the models.
    The differences in normalized fluxes between including and neglecting H+H
    collisions are smaller than 0.0005 of the continuum flux for the
    core-to-wing transition region that is most sensitive to $T_{\rm eff}$
    variations. 
    Therefore all the NLTE calculations were performed with pure electronic
    collisions with the exception of a very metal poor (VMP) giant atmosphere
    4600/1.60/$-2.6$ and a metal-poor subgiant atmosphere 5777/3.70/$-$2.38, for
    which S$_H$=1.0 and 2.0 were adopted.

    The coupled radiative transfer and SE equations were solved with a revised
    version of the code DETAIL \citep{Butler1985}.
    The update was described in \citet{Mashonkina2008}. 
    The obtained departure coefficients were then used by the code SIU
    \citep{Reetz1991} to calculate the synthetic line profiles.
    The absorption profiles of Balmer lines were computed by convolving the
    profiles resulting from the thermal, natural, and Stark broadening, as well
    as self-broadening.
    The Stark profile was based on the unified theory as developed by
    \citet[][VCS]{Vidal1970,Vidal1973}.
    The calculations of \citet{Stehle1994} based on a different theory agreed
    reasonably well with the data of VCS.
    For self-broadening, we applied the Lorentz profile with a half-width
    computed using the cross-section and velocity parameter from
    \cite{Barklem2000}.
    As shown by \citet{Barklem2000}, "the complete profiles obtained from
    overlapping line theory are very closely approximated by the p-d component
    of the relevant Balmer line".
    This gives ground to apply the impact approximation to describe the
    self-broadening effect on H$_\alpha$, H$_\beta$, and H$_\gamma$.

    \begin{figure}
    \resizebox{88mm}{!}{\includegraphics{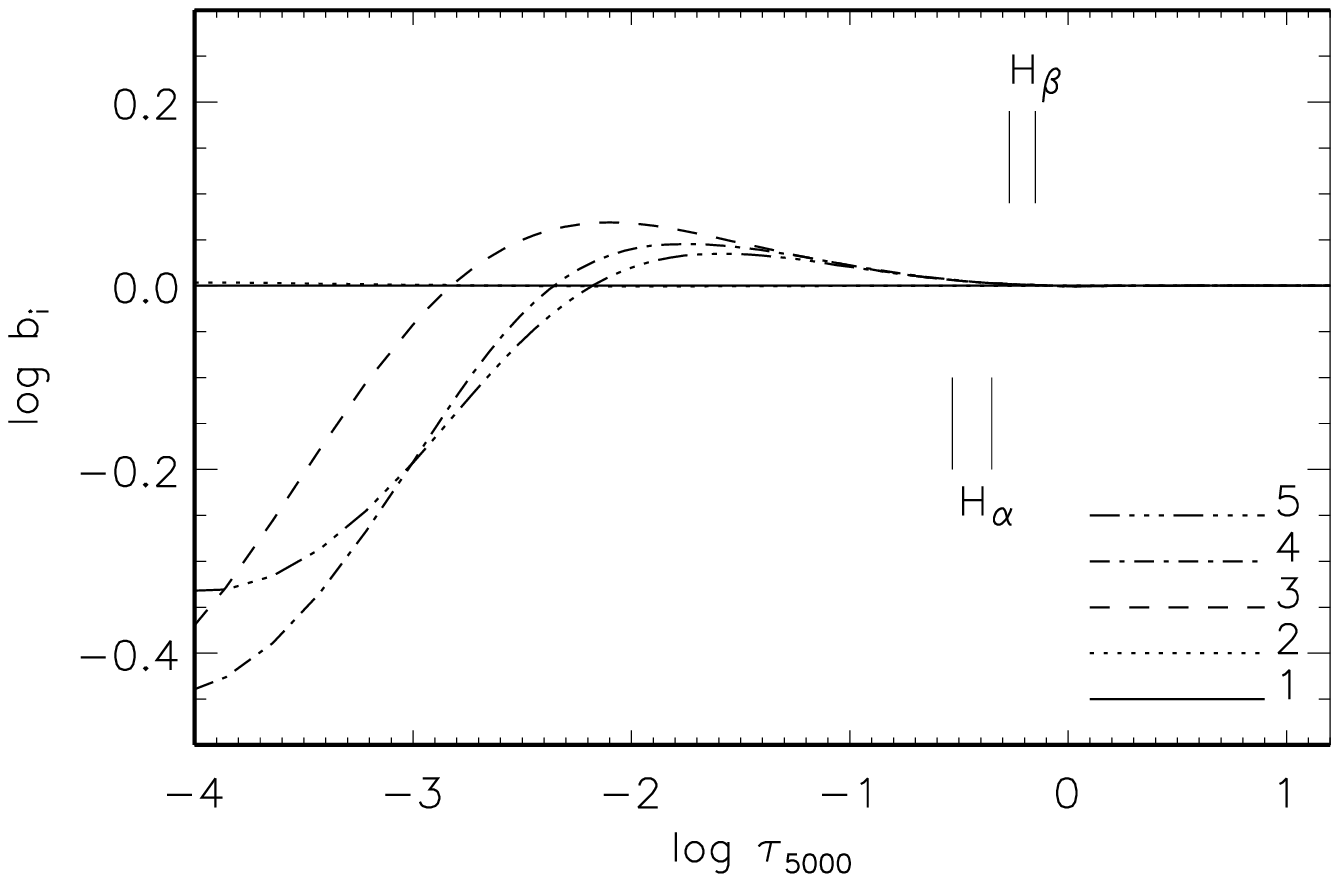}}
    \caption[]{Departure coefficients $\log b_i$ for the five lowest levels of
        \ion{H}{i} as a function of continuum optical depth $\tau_{5000}$
        referring to $\lambda = 5000$\AA\, in the model atmosphere
        6530/3.96/$-0.05$.
        Tick marks indicate the locations of core-to-wing transition (0.85 to
        0.95 in normalized flux) formation depths for H$_\alpha$ and H$_\beta$.}
    \label{h_bf}
    \end{figure}

\begin{figure}
    \resizebox{88mm}{!}{\includegraphics{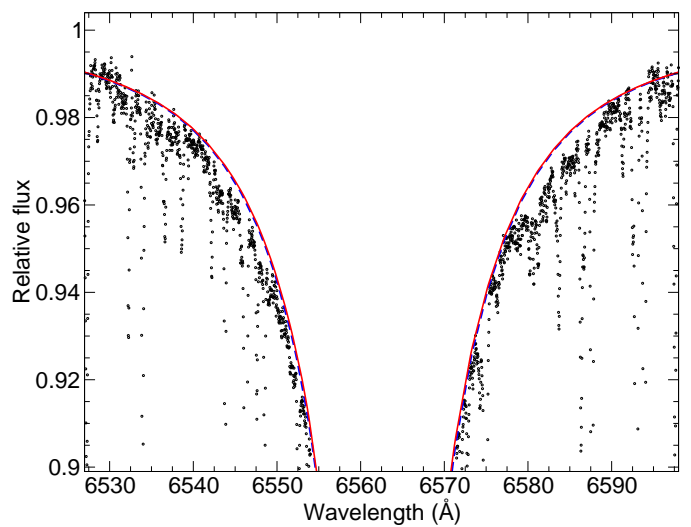}}
    \caption[]{
        Synthetic NLTE (continuous curve) and LTE (dashed curve) flux profile of
        H$_\alpha$ computed with the MAFAGS-OS 6530/3.96/$-0.05$
        ($\alpha_{\rm CMA}=2.3$) model compared to the observed spectrum of
        Procyon (bold dots).}
    \label{halpha_profiles}
    \end{figure}

    We selected the model 6530/3.96/$-0.05$ to show the behavior of the
    departure coefficients $b_i = n_i^{\rm NLTE}/n_i^{\rm LTE}$ of the relevant
    levels of \ion{H}{i} (Fig. \ref{h_bf}). Here, $n_i^{\rm NLTE}$ and
    $n_i^{\rm LTE}$ are the SE and thermal (Saha-Boltzmann) number densities,
    respectively.
    Very similar behavior of $b_i$ was found for all the model atmospheres
    investigated in this study. The ground state and the $n=2$ level keep
    their thermodynamic equilibrium level populations throughout the atmosphere.
    Departures from LTE for the $n=3$ level are controlled by H$_\alpha$.
    In the layers where the continuum optical depth drops below unity, H$_\alpha$
    serves as the pumping transition, which results in an overpopulation of the
    upper level.
    For instance, $b(n=3)\simeq1.017$ around $\log\tau_{5000}=-0.57$, where the
    H$_\alpha$ core-to-wing transition is formed.
    Starting from $\log\tau_{5000}\simeq-2.8$ and farther out, photons that
    escape from the H$_\alpha$ core cause an underpopulation of the $n=3$ level.
    For H$_\alpha$, NLTE leads to a weakening of the core-to-wing transition
    compared to the LTE case (Fig.\,\ref{halpha_profiles}).

    To compare them with observations, the computed synthetic profiles were
    convolved with a profile that combines instrumental broadening with a
    Gaussian profile of 3.2\,km\,s$^{-1}$ to 4.6\,km\,s$^{-1}$ for different
    stars and broadening caused by macroturbulence with a radial-tangential
    profile of 4\,km\,s$^{-1}$.
    Rotational broadening was only taken into account for Procyon, with $V\sin
    i$ = 2.6\,km\,s$^{-1}$ \citep{Fuhrmann1998}.
    All the remaining dwarf and subgiant stars were assumed to be slow rotators,
    with $V\sin i\le1$\,km\,s$^{-1}$.
    It is worth noting that the thermal broadening of Balmer lines, with the most
    probable velocity of about $V_t=10$\,km\,s$^{-1}$ in the atmospheres of the
    investigated stars, is more pronounced than the broadening caused by rotation,
    macroturbulence, and instrumental profile together.
    Therefore, the uncertainty in the external broadening parameters only has an
    minor effect on the computed Balmer-line wings.

\section{Fitting procedure of Balmer-line wings}

    Our analysis of H$_\alpha$ is based on profile fitting in the wavelength
    range from 6473 to 6653 \AA\, for Procyon and from 6520 to 6608 \AA\ for the
    remaining stars.
    We analyzed H$_\beta$ in the wavelength range 4815 -- 4908 \AA\ only for the
    metal-poor stars HD\,103095, HD\,122563, HD\,45282, and HD\,140283. 
    For HD\,122563 and HD\,140283 we also analyzed H$_\gamma$ profiles in the
    wavelength range from 4315 to 4365 \AA\ .
    Before fitting, we made the following preparations of the observed profiles
    for all stars.

    We excluded the Balmer line cores up to 0.9 of the relative flux
    (F$_{i}^l$/F$_{i}^c$), because the core half-width of the hydrogen line cannot
    be described in cool stars with classical model atmospheres.
    Thus only the pressure-broadened wings were analyzed.
    As our aim is to fit the profiles of Balmer lines, all blended lines should
    be removed from the observed profiles.
    The high resolution of the employed spectra allows us to assume that
    unblended spectral windows do exist and can be used for the fitting.
    We identified spectral windows across the line that are expected to be free
    from blends in the stars of interest. 
    This was done by means of calculating Balmer line profiles with and without
    metal lines.
    Then we created a mask that was free from any blends. 
    Based on the examination of the spectra, this is reasonable at H$_\alpha$
    for all the reference stars, and at H$_\beta$ for HD\,122563, HD\,45282, and
    HD\,140283, and at H$_\gamma$ only for HD\,122563 and HD\,140283.
    It is impracticable to define spectral windows without any blends at H$_\beta$
    and H$_\gamma$ in solar metallicity stars.

    The mask consists of a spectral window of pure H$_\alpha$, H$_\beta$, and
    H$_\gamma$ (when available) lines with noise alone.
    The same mask was applied for all stars, but minor modifications were made
    for each star individually.
    Modifications of the mask were necessary to reject stars in some specific
    cases, for example, if there was an ingress of atmospheric lines in the
    spectral window or for Procyon, which has much stronger macroturbulence and
    $V\sin i$ than other stars.
    It is very important to clearly distinguish between noise fluctuations and
    weak metal lines in the spectra.
    In this context, we relied upon the spectral line database and inspected all
    Balmer line profiles.

    We also investigated the effect of spectral resolution on the choice of
    which spectral windows to mask.
    It is necessary to have a proper number of spectral windows between blended
    lines because only by using such a mask are we able to derive the same result
    regardless of whether we use high-resolution spectra or the spectra degraded
    to the resolution of our observations.
    This mask is quite reliable in the case of H$_{\alpha}$ for all stars, but
    for H$_{\beta}$ and H$_{\gamma}$ we were unable to define such a mask for all
    stars.
    This makes the results obtained from H$_{\beta}$ and H$_{\gamma}$ profiles
    less reliable.

    We also took into account the effect of rotational broadening and broadening
    caused by macroturbulence on the choice of spectral windows of mask.
    While all our sample stars are slow rotators with $V\sin i\le1$\,km\,s$^{-1}$
    except for Procyon with $V\sin i=2.6$\,km\,s$^{-1}$, we were able to find a
    proper number of spectral windows between blending lines around H$_{\alpha}$
    profiles.
    In the case of H$_{\beta}$ and H$_{\gamma}$ profiles, this effect becomes
    important because there are many blending lines in the two profiles.
    This prevented us from defining a reliable mask for all stars, except for
    some metal-poor stars (HD\,103095, HD\,122563, HD\,45282, and HD\,140283),
    which only contain few blending lines in their H$_{\beta}$ and H$_{\gamma}$
    profiles.

    In this work we used a method based on a reduced $\chi^2$ statistics.
    Statistical measurement of the goodness-of-fit was performed by comparing
    the theoretical profile and an observed profile through a $\chi^2$
    minimization procedure \citep{Nousek1989}.
    The estimator of the fit goodness is defined by

    \begin{center}
    \begin{equation}
    \chi^2_{\rm red}=\frac{1}{n-p-1}\sum_{i=1}^{n} \frac{(O_i-C_i)^2}{\sigma_i^2}
    \label{chi}
    ,\end{equation}
    \end{center}

    \noindent where $O_i$ is the relative flux measured at the wavelength
    $\lambda_i$, $C_i$ is the relative flux computed from the model at the same
    wavelength $\lambda_i$, $\sigma_i$ is the variance of the data point $i$
    that is defined as the (S/N)$^{-1}$ ratio at the same wavelength
    $\lambda_i$, $n$ is the number of data points, and $p$ is the number of free
    parameters of the model so that $n-p-1$ is the number of degrees of freedom
    (d.o.f). 

    For the spectra of HD\,10700 and HD\,122563, we did not have the $\sigma_i$
    at the wavelength $\lambda_i$.
    Although it is known that the S/N depends on the number of recorded photons
    $I$ ($\sim$ $\sqrt{I}$), we adopted the mean constant S/N ratio for these
    stars because we only fit the wings of each line up to 0.9 where the S/N
    ratio is not changed dramatically.
    For Procyon, for example, the relative flux at a S/N of 0.85 is 10\%\ lower
    than that with a S/N of 0.995.
    Apart from the inner wings, which have less strongly blended lines because
    they are saturated by the hydrogen line, equal weights were assigned for all
    spectral windows of a line profile.

    There were two free parameters, $\alpha_{\rm CMA}$ and the effective
    temperature, while the other parameters ($\log{g}$, [Fe/H], $\xi$) were
    fixed.
    Assuming the Poissonian error distribution, which approaches the normal
    distribution at high S/N, the minimization of $\chi^2$ provides the
    maximum-likelihood estimate of the parameters, in our case for
    $\alpha_{\rm CMA}$ and $T_{\rm eff}$.

\begin{figure}
    \resizebox{88mm}{!}{\includegraphics{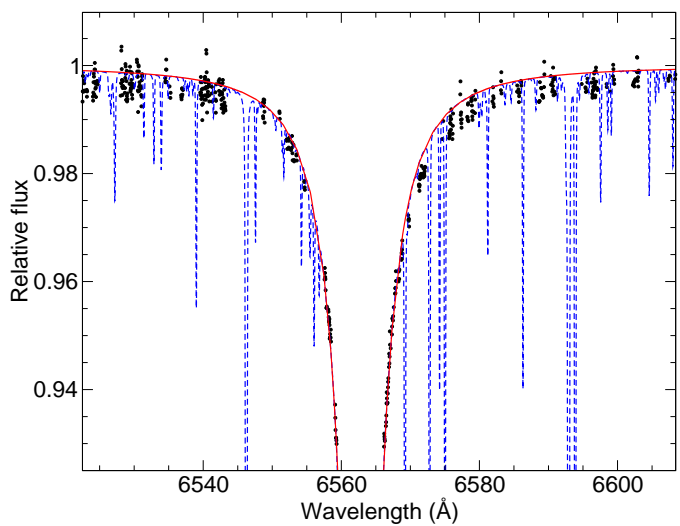}}
    \caption[]{Example of fitting the H$_\alpha$ wings in the metal-poor star
        HD\,103095.         
        Bold dots correspond to the observed data.
        The theoretical H$_\alpha$ profile was calculated with (dashed curve) and
        without (solid curve) blending lines. Only bold dots were used for        determining the $\chi^2$ statistic.}
    \label{example}
    \end{figure}

    We also compared a grid of Balmer-line profiles for a given model grid with
    a 25 K interval in $T_{\rm eff}$ and a 0.1 interval in $\alpha_{\rm CMA}$.
    By varying the free parameters of the model, we computed the value of $\chi^2$
    at each step.
    First we fixed one of the two free parameters, $T_{\rm eff}$, and analyzed
    the function of $\chi^2$, which smoothly changes with varying $\alpha_{\rm CMA}$
    and follows an asymmetric parabolic form with one single minimum.
    The best $\alpha_{\rm CMA}$ and $T_{\rm eff}$ were chosen from the requirement
    of the lowest $\chi^2$ value. 
    The variation of temperature within the error bars allows us to estimate the
    upper and lower values for our $\alpha_{\rm CMA}$.
    An example of the fitting for H$_\alpha$ profile in the metal poor star
    HD\,103095 is shown in Fig.\ref{example}.

    \subsection{Errors}

    Estimated errors are presented in Table \ref{tab2} for the program stars.
    The strength of the hydrogen line depends only weakly on the surface gravity
    or the metallicity.
    We considered how the typical errors (0.1 dex) for either of them may affect
    the result.
    Microturbulence does not affect the formation of the Balmer-line wings
    \citep{Fuhrmann1993}, therefore we did not consider the influence of
    microturbulence.
    Only the temperature can strongly affect the wings of hydrogen lines.
    With increasing temperature the wings of hydrogen lines tend to be stronger.
    The effective temperature is the main source of errors in our results. 
    The influence of other sources of error, such as Stark-broadening,
    self-broadening, and He-broadening, were considered in \citet{Fuhrmann1993,
    Fuhrmann1994} and \citet{Barklem2002}.
    The latter showed that the influence of Stark- and self-broadening on the
    behavior of Balmer lines is insignificant for solar metallicity stars, but
    tends to be stronger for metal-poor atmospheres.
    Moreover, the determination of the continuum placement is one of the most
    significant uncertainties in this work.
    For the sample spectra, the error in the continuum placement of the
    high-quality spectra is approximately 0.5\% of the continuum flux, depending
    on the S/N ratio. 
    As a result of the interdependency of errors, it is difficult to precisely
    account for the combined effect, therefore we divided all errors into two
    groups: those propagated from the stellar parameters ($T_{\rm eff}$,
    $\log{g}$, and [Fe/H]), and those caused by observational aspects (continuum
    placement and fitting error).

\begin{table*}
 \begin{minipage}{170mm} 
 \renewcommand{\arraystretch}{1.5}
 \tiny{
 \caption[]{Estimated errors in CMA factor for H$_\alpha$ profiles. }
\label{tab2}
\begin{tabular}{cccccccccccc} \\ \hline 
Error &  \tiny{Sun}  &  \tiny{Procyon} &  \tiny{HD\,10700} & \tiny{HD\,103095} &
\tiny{HD\,39587} & \tiny{HD\,6582} & \tiny{HD\,217014} & \tiny{HD\,22049} &
\tiny{HD\,122563} & \tiny{HD\,45282} & \tiny{HD\,140283}   \\ \hline   
$\Delta T_{\rm eff}$      &$\pm0.2$ &$\pm0.4$ &$\pm0.3$ &$\pm0.3$ &$\pm0.4$ &$\pm0.3$ &$\pm0.3$ &$\pm0.3$ &$\pm0.1$ &$\pm0.3$ &$\pm0.2$\\ 
$\Delta\log{g}$ = 0.1 dex & --      &--       &--       &$\pm0.1$ &--       &$\pm0.1$ &--       & --      &$\pm0.1$ &$\pm0.1$ &$\pm0.1$\\ 
$\Delta$[Fe/H] = 0.1 dex  & --      &$\pm0.2$ &$\pm0.1$ &$\pm0.1$ &--       &$\pm0.1$ &$\pm0.1$ &$\pm0.1$ &$\pm0.1$ &--       &--      \\   
Fitting procedure         &$\pm0.2$ &$\pm0.2$ &$\pm0.1$ &$\pm0.2$ &$\pm0.1$ &$\pm0.5$ &$\pm0.1$ &$\pm0.1$ &$\pm0.2$ &$\pm0.2$ &$\pm0.2$\\ 
Continuum                 &         &         &         &         &         &         &         &         &         &         &        \\ 
          (0.5\%)         &$\pm0.3$ &$\pm0.5$ &$\pm0.2$ &$\pm0.3$ &$\pm0.1$ &$\pm0.3$ &$\pm0.1$ &$\pm0.1$ &$\pm0.2$ &$\pm0.2$ &$\pm1.0$\\ \hline 
Total                     &$\pm0.4$ &$\pm0.7$ &$\pm0.4$ &$\pm0.5$ &$\pm0.4$ &$\pm0.7$ &$\pm0.4$ &$\pm0.4$ &$\pm0.4$ &$\pm0.4$ &$\pm1.1$\\ \hline  
 \end{tabular}
 }
 \end{minipage}
\end{table*}

\section{Results}

    \begin{figure*}[ht]
        \centering
        \includegraphics[width=18cm]{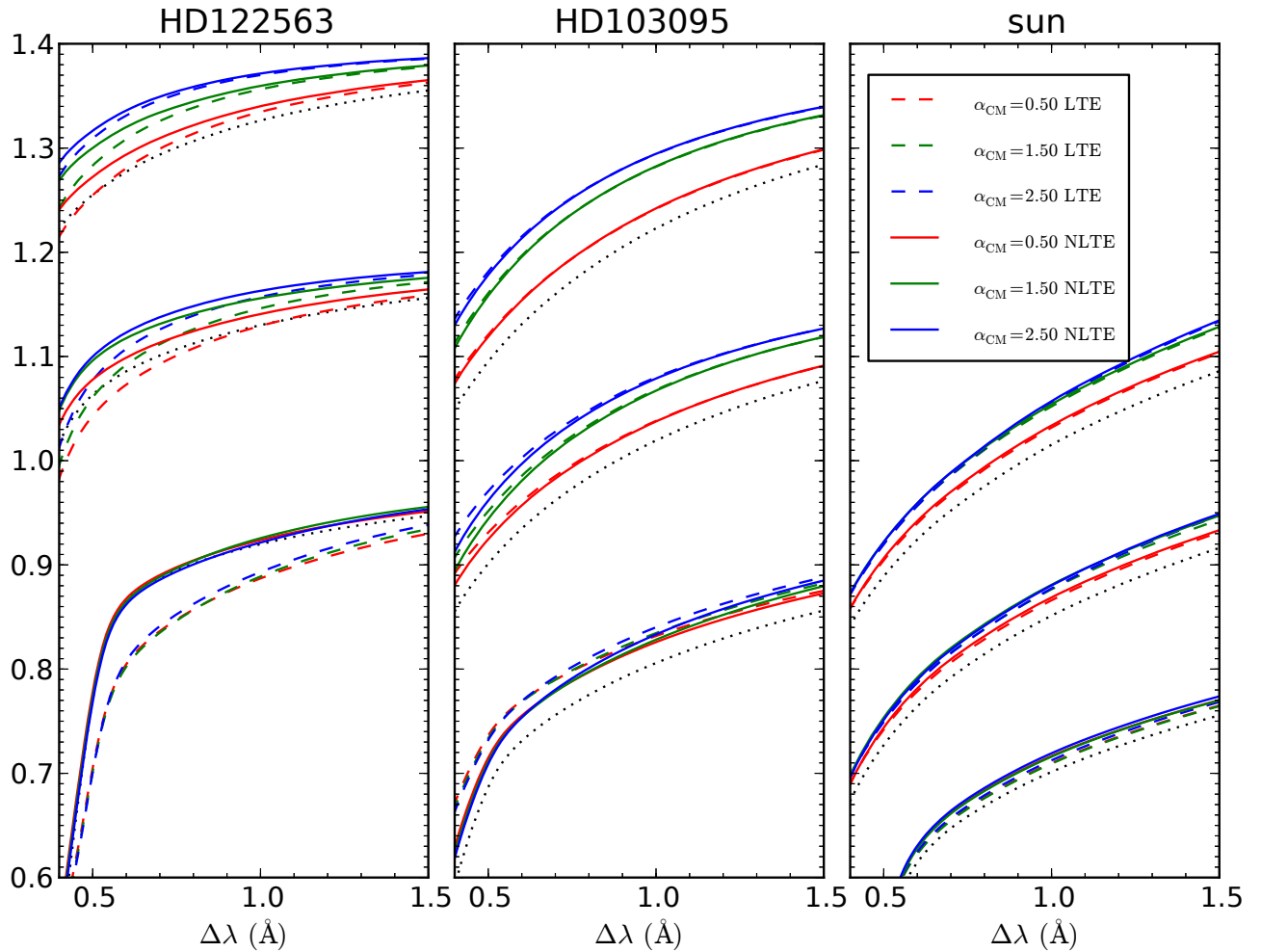}
        \caption{
            Comparison of the line profiles of H$_\alpha$, H$_\beta$, and
            H$_\gamma$ (from bottom to top) for HD\,122563, HD\,103095, and the
            Sun.
            The x-axis is the wavelength shift relative to the center of the
            Balmer lines.
            The colors are coded with different $\alpha_{\rm CMA}$ (from 0.5 to
            1.5),
            while the solid lines and dashed lines represent profiles using LTE
            and NLTE line formations, respectively.
            The dotted lines are profiles with LTE models of $\alpha_{\rm CMA}
            =0.5$ but $T_{\rm eff}+100$ K.
            For clarity, offsets of 0.2 and 0.4 are added to the H$_\beta$ and
            H$_\gamma$, respectively.
        }
        \label{fig-profiles}
    \end{figure*}

    In Fig. \ref{fig-profiles} we plot the profiles of the Balmer lines for
    three of our program stars (HD\,122563, HD\,103095, and the Sun).
    The profiles are only slightly different under LTE or NLTE assumptions for
    dwarfs, but exhibit significant variations with $\alpha_{\rm CMA}$ on their
    wings.
    The changes with $\alpha_{\rm CMA}$ from 0.5 to 2.5 are similar (for
    H$_\alpha$) and even larger (for H$_\beta$ and H$_\gamma$) to that of
    $T_{\rm eff}$ of a few hundreds of Kelvin.
    For the metal-poor giant HD\,122563, the NLTE effect plays a more important
    role than $\alpha_{\rm CMA}$ for H$_\alpha$ profile, but is less sensitive
    than $\alpha_{\rm CMA}$ parameter.
    Figure \ref{fig-profiles} also indicates that if the $\alpha_{\rm CMA}$
    parameter in the model atmosphere is not correct, the $T_{\rm eff}$ may be
    biased by up to a few hundreds of Kelvin by fitting H$_\beta$ or H$_\gamma$
    line profiles.
    Therefore, caution must be exercised when choosing $\alpha_{\rm CMA}$ in 1D
    model atmospheres.

    \subsection{The Sun}

    The Sun is the best-known star with high-quality observations and accurate
    physical parameters.
    Following \citet{Fuhrmann1993}, we attempted to calibrate $\alpha_{\rm CMA}$
    for the solar atmosphere.
    Our work differs from that of \citet{Fuhrmann1993} in that we have used the
    self-broadening theory of \citet{Barklem2000}.
    For investigations of the Sun we employed the solar atlas \citep{Kurucz1984}. 
    We adopted an effective temperature of 5777~K with an error bar of $\pm$40~K.
    By fitting theoretical H$_\alpha$ profiles to reduced observational profiles,
    the two parameters $T_{\rm eff}$ and $\alpha_{\rm CMA}$ were varied within
    error bars and reasonable physical limits, respectively.
    The results are shown in Fig. \ref{Sun}, where the $\chi^2$ minimum
    indicates the best $T_{\rm eff}$ and $\alpha_{\rm CMA}$.  

    Figure \ref{Sun_Proc_Profile} compares the left wing of H$_\alpha$ line
    profile for the Sun with the theoretical spectrum for $\alpha_{\rm CMA}=2.0$
    (solid line). 

    The combination of $T_{\rm eff}$=5770$\pm20$ K and $\alpha_{\rm CMA}=2.0\pm
    0.2$ is preferred based on the degree of agreement between the observational
    and theoretical H$_\alpha$ profiles.
    Nevertheless, it is important to note that the value of $\chi^2$ is still
    high, which suggests that our agreement between the observational and the
    theoretical H$_\alpha$ profiles is not perfect.
    This means that the quality of the observed solar flux spectra is higher
    than the quality of theoretical surveys. 

    From comparison with SED, a low value for $\alpha_{\rm CMA}<1.0$ is excluded.
    The lack of fitting in the region below 5000\,\AA\ does not allow precisely
    determining the convective efficiency, but the measurements are consistent
    with an $\alpha_{\rm CMA}$ of 2.0.

    \subsection{Procyon}

    Procyon (HD\,61421, HR\,2943) is one of the brightest stars in the night sky
    ($V=0.34$).
    It is a binary system consisting of an F5\,IV-V star (Procyon\,A) with a
    mass of $1.497\pm0.037\ M_\odot$ and a faint white dwarf (Procyon\,B) with a
    mass of $0.602\pm0.015\ M_\odot$, orbiting each other with a period of 40.8
    years \citep{Girard2000}.
    Spectroscopic analyses gave rather discrepant effective temperatures, from
    6470 K (by fitting the Balmer lines; \citealt{Zhao2000}) to 6850 K (by
    measuring the equivalent widths of a large set of iron lines;
    \citealt{Heiter2003}).
    The proximity of Procyon ($\pi=284.56\pm1.26$ mas; \citealt{vanLeeuwen2007})
    to the Earth allows us to directly measure its angular diameter
    \citep[e.g.,][]{Hanbury1974,Mozurkewich1991,Aufdenberg2005,Chiavassa2012}.
    \citet{Kervella2004} found $T_{\rm eff}=6530\pm50$ K and $\log{g}=3.96\pm0.2$,
    which agrees well with other works \citep[e.g.,][]{AP2002,Fuhrmann1997}.
    We used a high-quality spectrum of H$_\alpha$ for Procyon and applied the
    same fitting method as described in the previous sections.

\begin{figure}[ht]
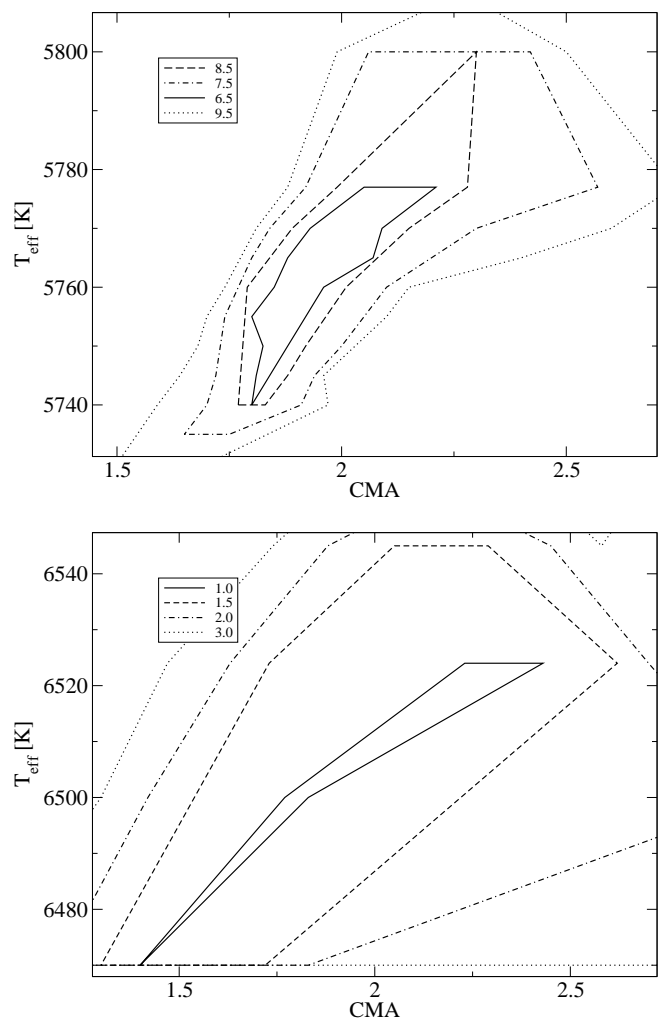

    \begin{center}
    \parbox{1\linewidth}{\includegraphics[width=8.0cm,trim=15mm 0mm 1mm 0mm]{3D_Sun}\\
    \centering}
    \hspace{1\linewidth}
    \hfill
    \parbox{1\linewidth}{\includegraphics[width=8.0cm,trim=15mm 0mm 1mm 0mm]{3D_Procyon}\\
    \centering}
    \\[-1cm]
    \end{center}\vspace{-5mm}
    \caption{Survey of the reduced $\chi^2$ obtained from H$_\alpha$ profile
        fitting for the observed solar flux spectra (top panel) and spectra of
        Procyon (bottom panel) with different $\alpha_{\rm CMA}$ and $T_{\rm
        eff}$.
        The preferred parameter area is located where $\chi^2$ is minimal. 
        All $\chi^2$ are not normalized and indicate a good fit in each case.}
    \label{Sun}
    \end{figure}

\begin{figure}[ht]
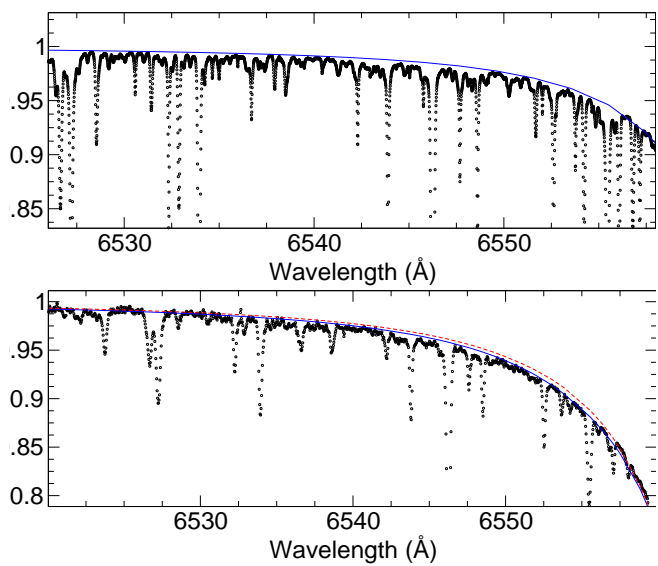

    \begin{center}
    \parbox{1\linewidth}{\includegraphics[width=8.5cm,clip=true,trim =15mm 0mm 1mm 0mm]{Halpha_SUN_20}\vspace{0mm}
    \centering}
    \parbox{1\linewidth}{\includegraphics[width=8.5cm,clip=true,trim =15mm 0mm 1mm 0mm]{Halpha_Proc_15_23}
    \centering}
    \\[0ex]
    \end{center}\vspace{-2mm}
    \caption{Left wings of H$_\alpha$ line profiles for the Sun with
        $\alpha_{\rm CMA}=2.0$ (top panel) and Procyon (bottom panel), where two
        theoretical spectra with $\alpha_{\rm CMA}=1.5$ (solid line) and
        $\alpha_{\rm CMA}=2.3$ (dash line) are shown.}
    \label{Sun_Proc_Profile}
    \end{figure}

    The results are displayed in Fig. \ref{Sun}, where the $\chi^2$ minimum
    indicates the best $T_{\rm eff}$ and $\alpha_{\rm CMA}$.
    For this star we find $T_{\rm eff}=6524\pm50$ K and $\alpha_{\rm CMA}=1.9\pm
    0.7$.
    The error of $\alpha_{\rm CMA}$ is not symmetric in this case, but we adopted
    the largest. 
    The left wing of the H$_\alpha$ line profile for Procyon is shown in Fig.
    \ref{Sun_Proc_Profile}, where two theoretical spectra with $\alpha_{\rm CMA}
    =1.5$ (solid line) and $\alpha_{\rm CMA}=2.3$ (dashed line) are plotted as
    well. 

    The value preferred from the SED fitting for $\alpha_{\rm CMA}\approx1.4$,
    nevertheless, there are large uncertainties with respect to opacity data.
    As a result of the uncertain UV flux comparison, the values given by SED
    fitting and that by Balmer profiles do not contradict each other.

    \subsection{Program stars}

    The final results of our fitting are given in Table \ref{table-params},
    including the names of the program stars, their parameters, the best
    $\alpha_{\rm CMA}$ along with the limits, and the $\chi^2$ values that
    indicate the best fit in each case.
    We also present the best-fit $T_{\rm eff}$ within the error bars that was
    obtained during the $\chi^2$ minimization procedure. 
    Temperatures with asterisks are the best-fit temperatures.
    The results from H$_{\beta}$ and H$_{\gamma}$ for some stars are less
    reliable.
    Even though good-quality observations of H$_{\beta}$ and H$_{\gamma}$ are
    available for all remaining stars, these lines were not adopted because
    choosing a reliable mask was problematic.

\begin{figure*}[h]
\begin{center}
\parbox{0.3\linewidth}{\includegraphics[scale=0.37]{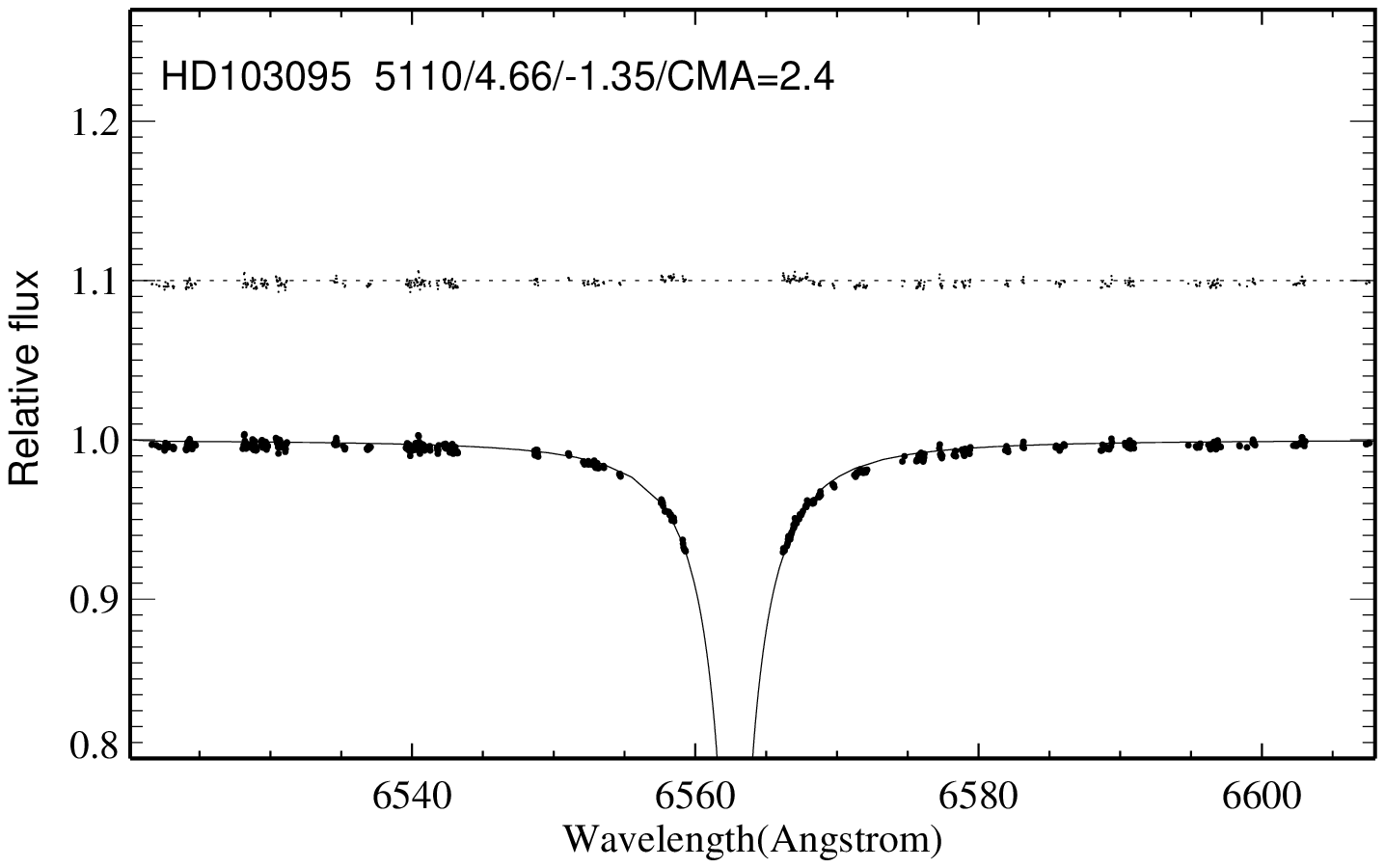}\\
\centering}
\parbox{0.3\linewidth}{\includegraphics[scale=0.37]{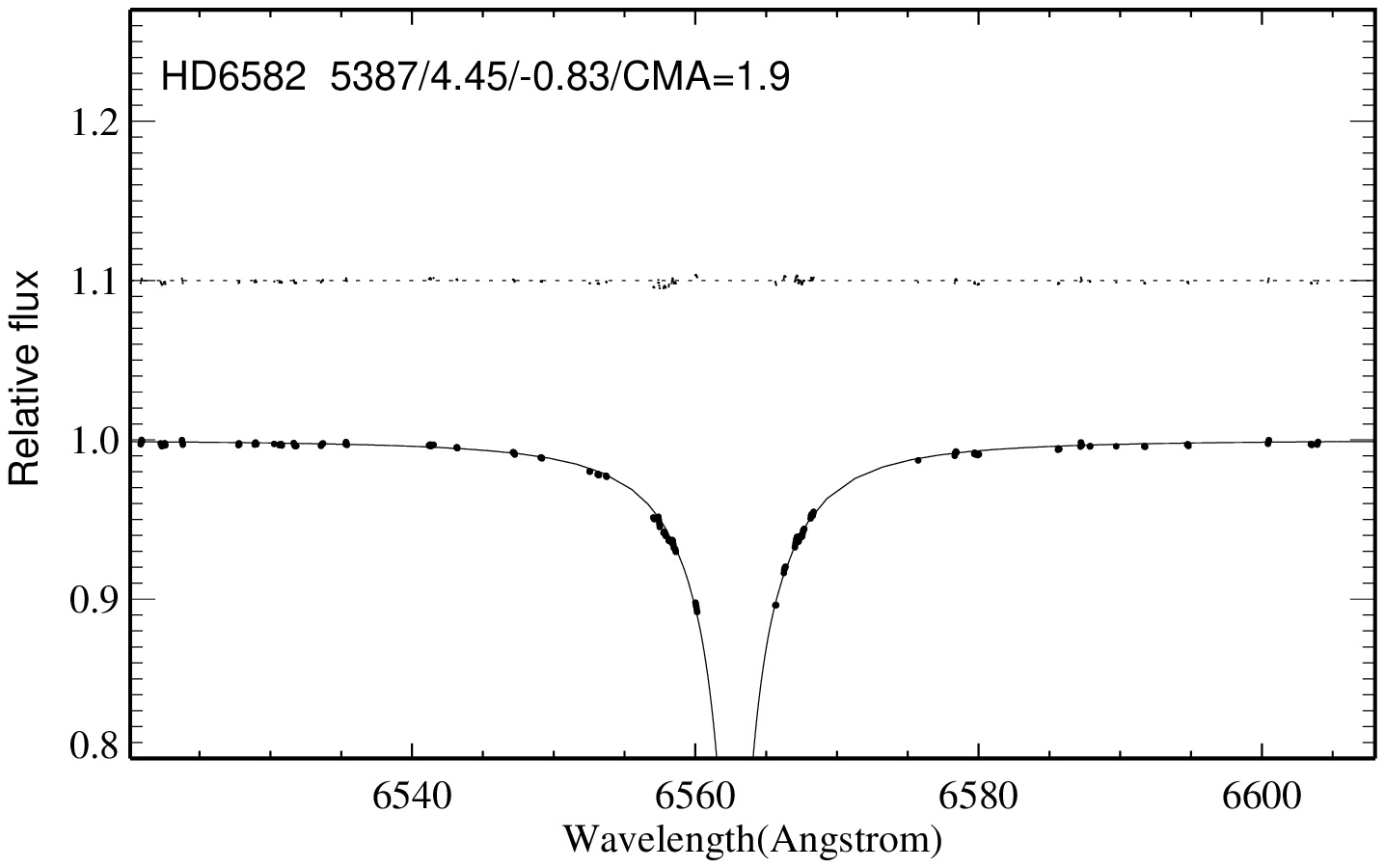}\\
\centering}
\parbox{0.3\linewidth}{\includegraphics[scale=0.37]{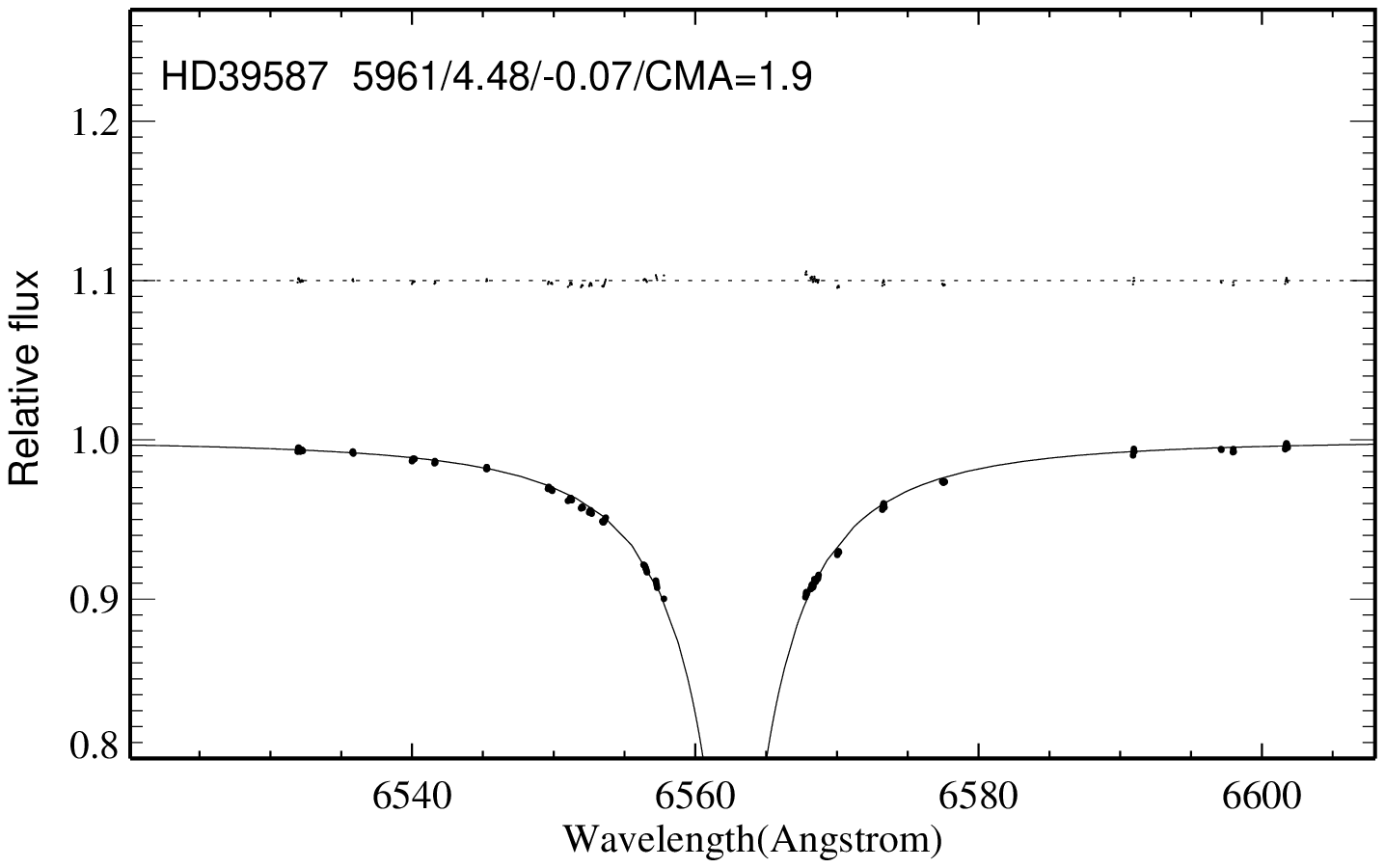}\\
\centering}
\hspace{1\linewidth}
\hfill
\\[0ex]
\parbox{0.3\linewidth}{\includegraphics[scale=0.37]{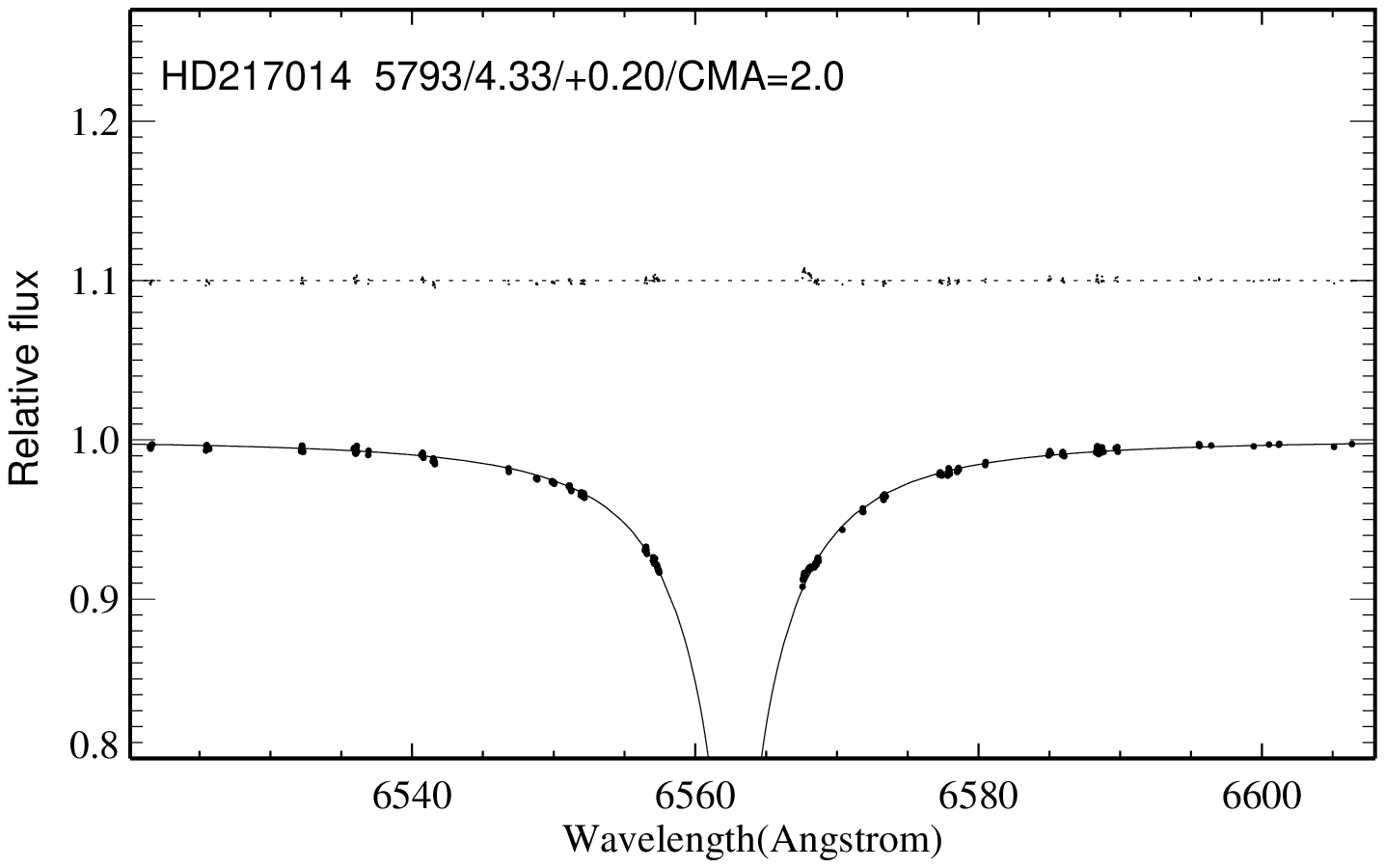}\\
\centering}
\parbox{0.3\linewidth}{\includegraphics[scale=0.37]{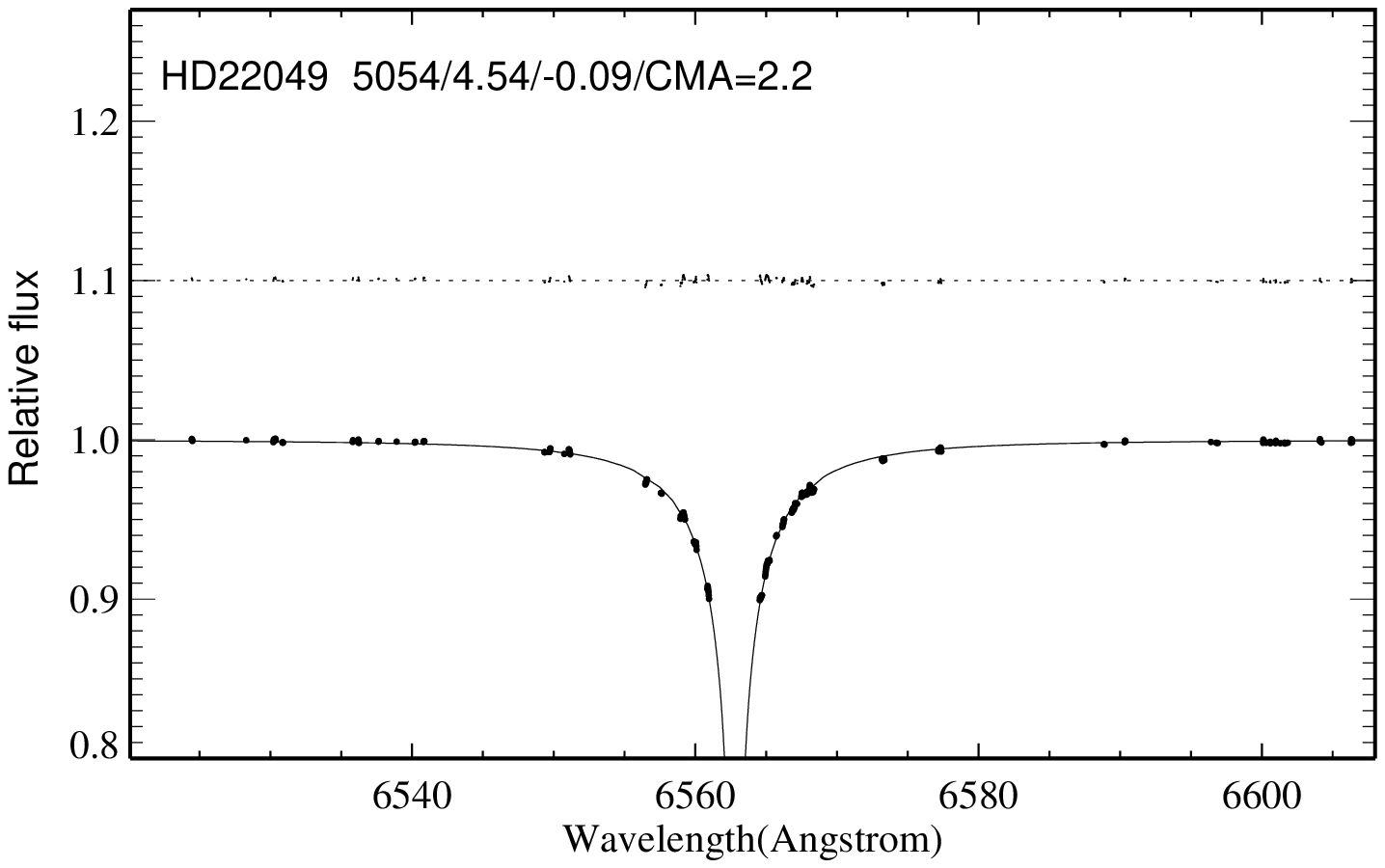}\\
\centering}
\parbox{0.3\linewidth}{\includegraphics[scale=0.37]{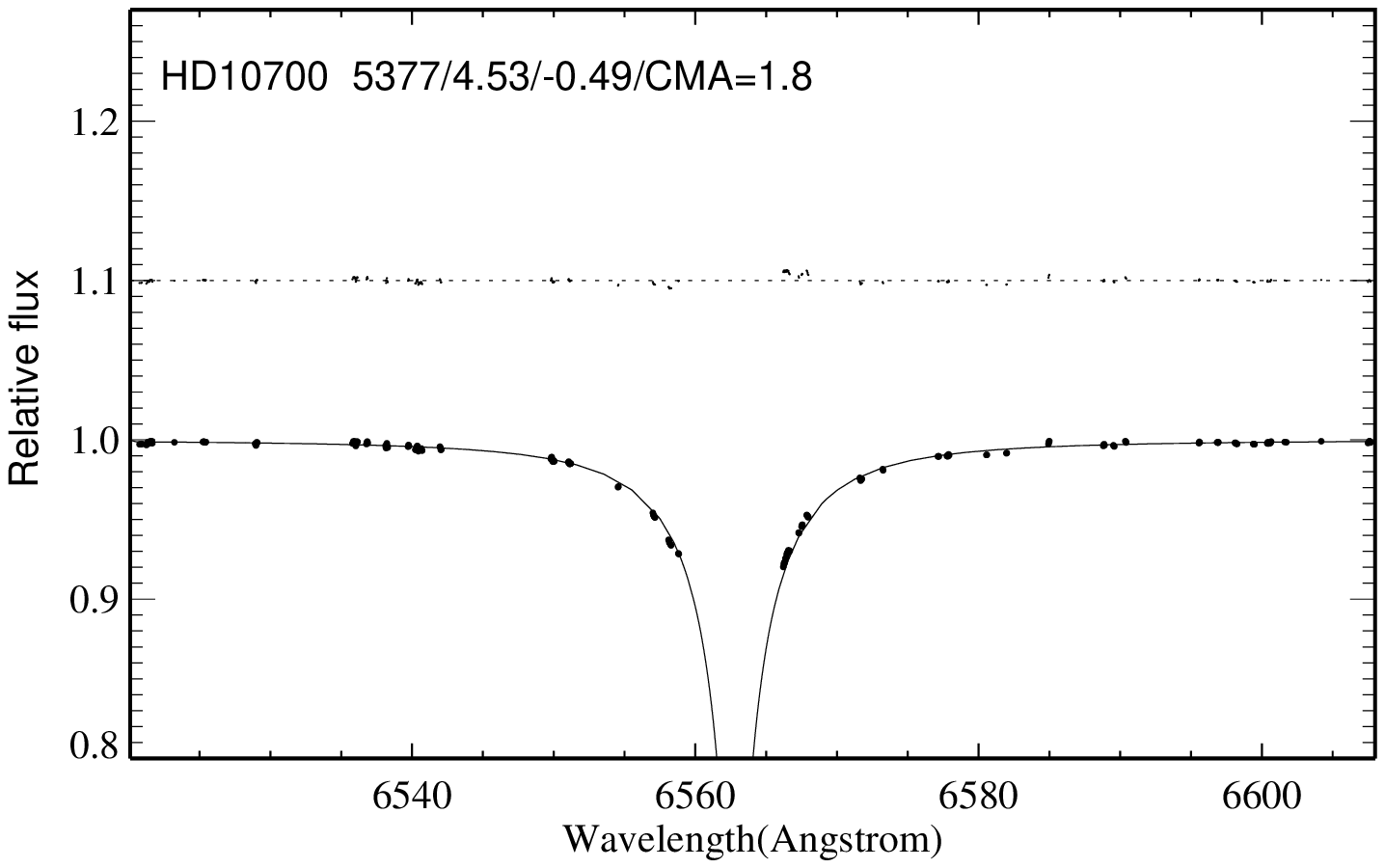}\\
\centering}
\hfill
\\[0ex]
\parbox{0.3\linewidth}{\includegraphics[scale=0.37]{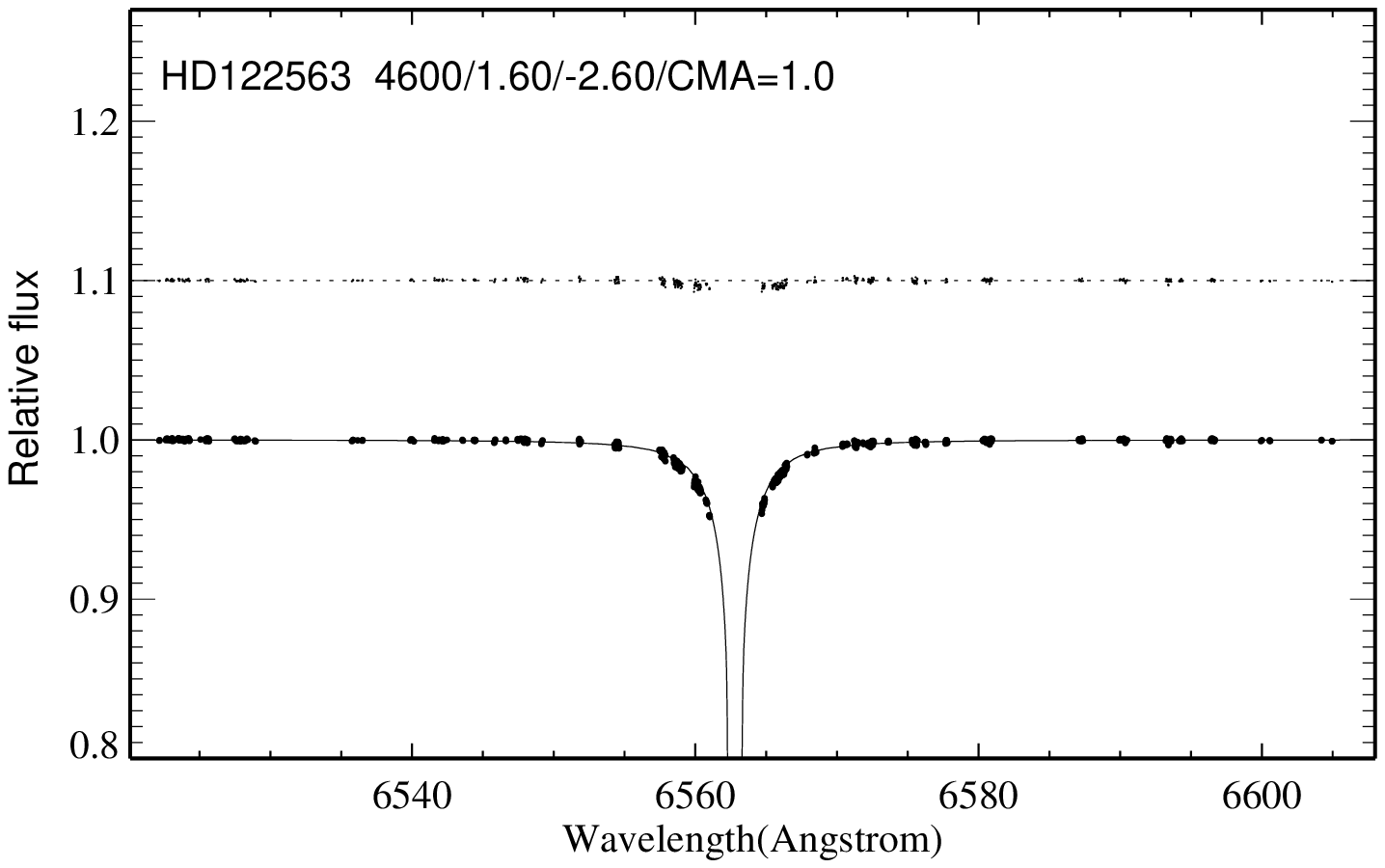}\\
\centering}
\parbox{0.3\linewidth}{\includegraphics[scale=0.37]{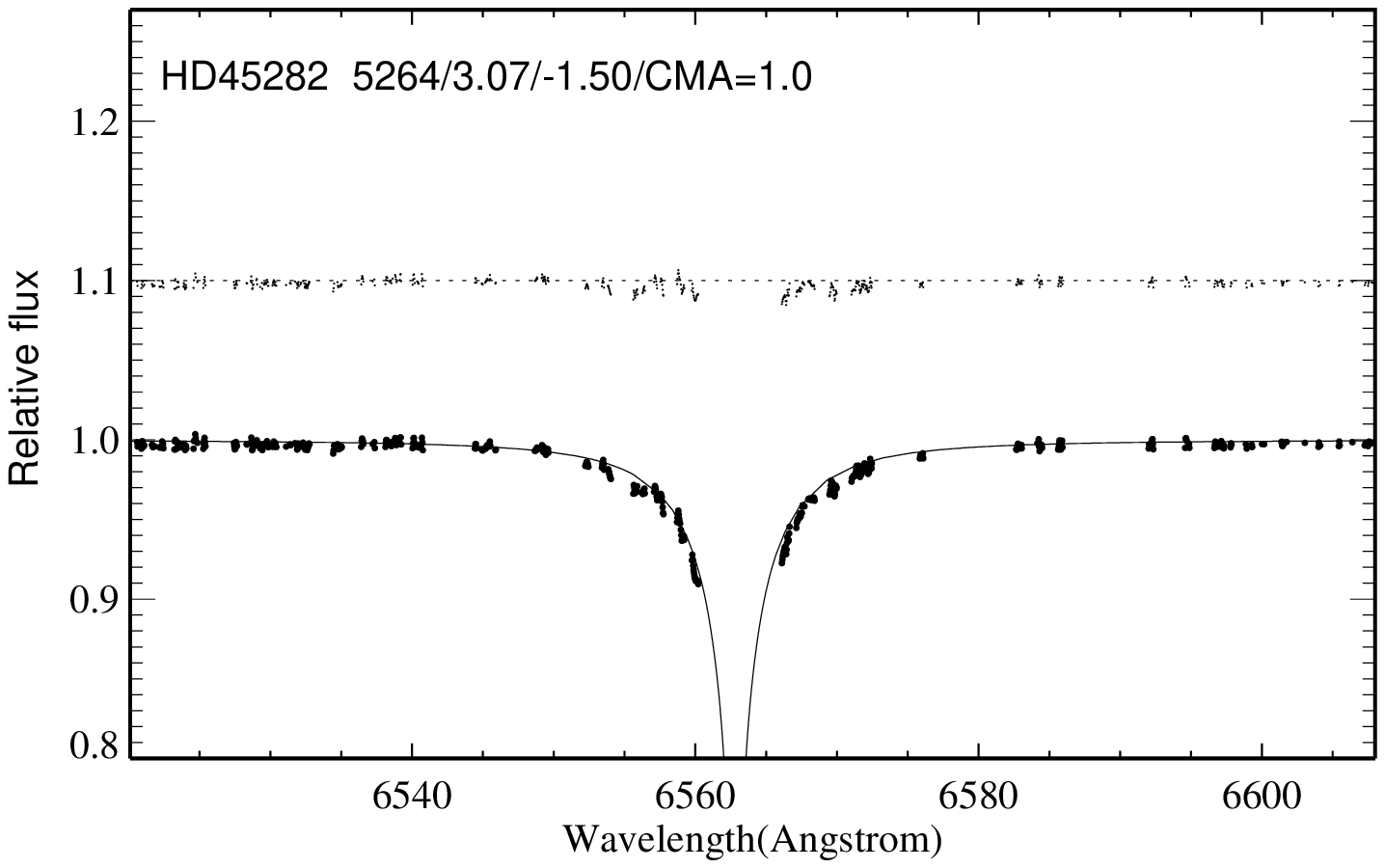}\\
\centering}
\parbox{0.3\linewidth}{\includegraphics[scale=0.37]{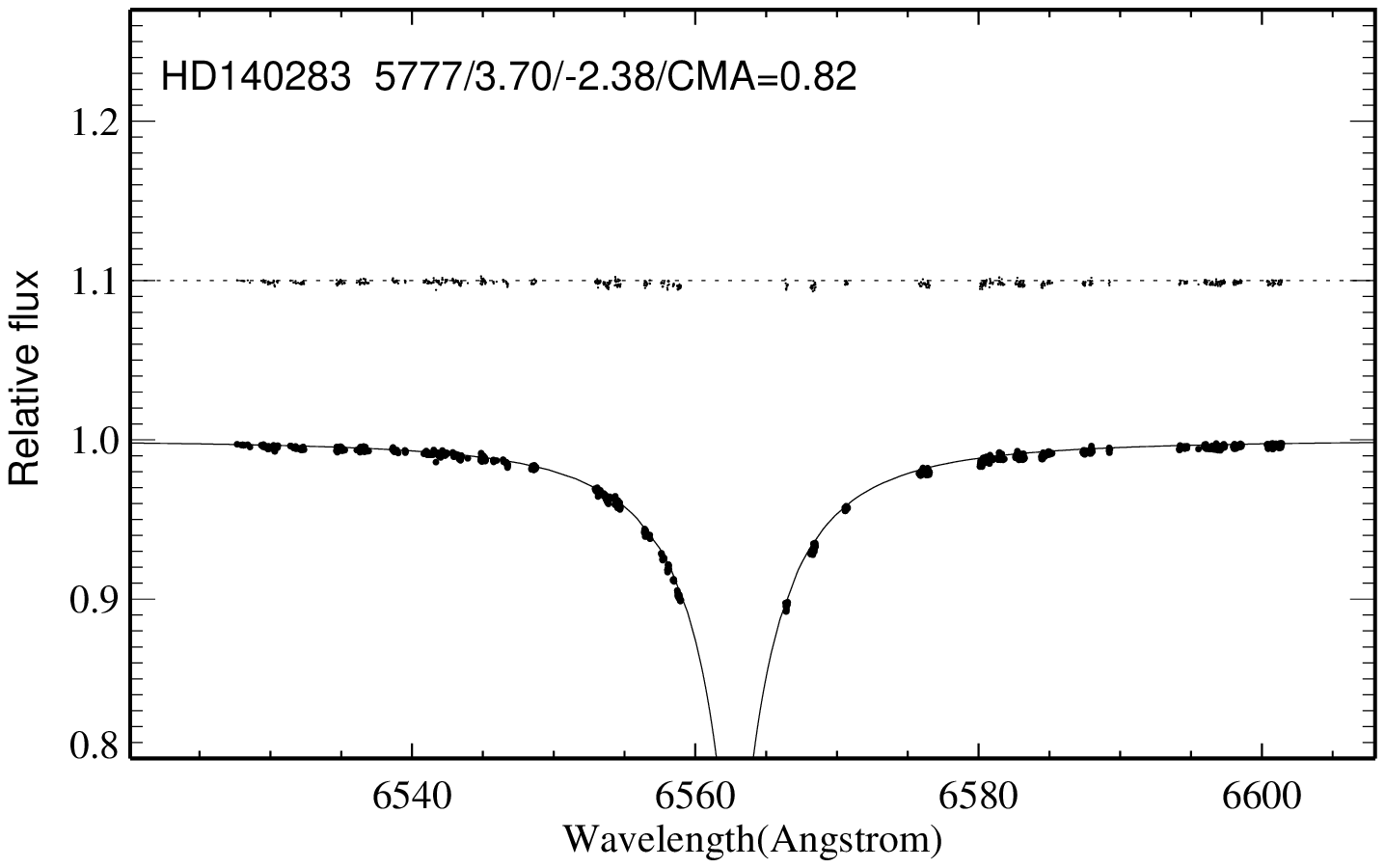}\\
\centering}
\hfill
\\[0ex]
\end{center}
\caption{Best fits of the H$_\alpha$ lines in our stellar sample.
    Solid lines are the theoretical spectra corresponding to the best fit and
    dotted lines represent the observed spectra that were used for determining
    the $\chi^2$ statistic.
    The horizontal dotted lines show the limit of reasonableness of the
    approximation and allows seeing residuals.}
\label{profiles}
\end{figure*}

    The observed spectra and the best-fitting line profiles from this method are
    plotted in Fig. \ref{profiles}. When applying the above-mentioned method, we
    have to keep in mind that weak blending lines may, nevertheless, be presented
    in a chosen windows.
    Therefore, it would be better to inspect the best profiles visually to give
    preference to the case where all observed points are below the theoretical
    line profile, but the $\chi^2$ -- method minimizes the deviations without
    considering whether the observed points are above or below the theoretical
    profile.
    In this context, for the $\chi^2$ method, $\alpha_{\rm CMA}$ will be slightly
    underestimated, while the best-fit $T_{\rm eff}$ will be overestimated. 

    The derived $T_{\rm eff}$ for stars HD\,10700, HD\,39587, HD\,217014,
    HD\,22049, HD\,122563, HD\,45282, and HD\,140283 agree well with those
    derived from astrometric and IRFM methods. 
    However, for HD\,103095 we were unable to find an appropriate $T_{\rm eff}$
    in the temperature range $T_{\rm eff}=4820\pm100$ K.
    Thus, for this star we adopted a higher spectroscopic temperature of
    $T_{\rm eff}=5110$ K.

    The individual results for our stellar sample are reported below.
 
    HD\,6582 ($\mu$\,Cas, 30\,Cas, HR\,321) is a close binary system that
    consists of a slightly metal-deficient subdwarf (component A, $V=5.17$) and
    a component B fainter by 5.3 magnitudes ($V=10.5$), with a separation of
    only 0".40$\sim$1".40 \citep{Mason2001}.
    The spectroscopic parameters adopted here for HD\,6582\,A ($T_{\rm eff}=
    5387$ K, $\log{g}=4.45$, [Fe/H] = $-0.83$, and $\xi=0.89$ km\,s$^{-1}$) were
    taken from \cite{Fuhrmann2004}.
    \cite{Boyajian2012a} measured its limb-darkened angular diameter
    $\theta_{\rm LD}=0.972\pm0.009$ mas with CHARA, corresponding to the stellar
    radius of $0.790\pm0.009\,R_\odot$, combining with the Hipparcos parallax
    ($\pi=132.38\pm0.82$ mas; \citealt{vanLeeuwen2007}).
    This is agrees well with the spectroscopic radius ($R=0.8\,R_\odot$), but
    the interferometric $T_{\rm eff}=5264$ K is $\sim$120 K lower than that of
    \cite{Fuhrmann2004}.
    By comparing the H$_\alpha$ profile, we found the best $\alpha_{\rm CMA}=
    2.3$ for spectroscopic parameters, while $\alpha_{\rm CMA}=1.7$ for
    interferometric parameters.
    Despite this large discrepancy, the fitting method with the interferometric
    model (b) led to lower $\chi^2$ and supports a higher $T_{\rm eff}=5370$ K.
    This result supports the relative high $T_{\rm eff}$ of 5339 K derived by
    the photometric $b-y=0.437$ and $c_1=0.213$ \citep{Hauck1998} with the
    color-$T_{\rm eff}$ relation of \citet{Alonso1996}.

    HD\,10700 ($\tau$\,Cet, 52\,Cet, HR\,509) is a nearby and inactive G8\,V
    star, with a distance of only $3.650\pm0.002$\,pc ($\pi=273.96\pm0.17$ mas;
    \citealt{vanLeeuwen2007}).
    It is one of the most frequently observed targets in several precise radial
    velocity surveys for extrasolar planets, and \cite{Tuomi2013} reported five
    planets with minimum masses of up to 6.6 $M_\oplus$ orbiting it.
    Spectroscopic analyses indicate that its $T_{\rm eff}$ ranges from 5283 K
    \citep{Valenti2005} to 5420 K \citep{Takeda2005}, and [Fe/H] range from
    $-$0.43 \citep{Takeda2005} to $-$0.58 \citep{Luck2005}.
    Interferometric measurements with VLTI gave $\theta_{\rm LD}=2.078\pm0.031$
    mas \citep{DiFolco2004}, and $T_{\rm eff}$ was perfectly consistent with the
    spectroscopic parameters ($T_{\rm eff}=5377$ K, $\log{g}=4.53$, [Fe/H] =
    $-0.49$, and $\xi=0.8$ km\,s$^{-1}$; \citealt{Mashonkina2011}).
    With a least $\chi^2$ fitting, we found the best $\alpha_{\rm CMA}=1.8$, and
    the best $T_{\rm eff}=5330$ K agrees with the $T_{\rm eff}=5346$ K inferred
    from the $b-y$ and $c_1$ color \citep{Hauck1998,Alonso1996}.

    HD\,22049 ($\epsilon$\,Eri, 18\,Eri, HR\,1084) is a nearby ($\pi=310.94 \pm
    0.16$ mas; \citealt{vanLeeuwen2007}), chromospherically active K2\,V star
    orbited by a dusty ring \citep{Greaves1998}.
    Atmospheric parameter determinations based on high-resolution spectra gave
    very consistent results, with a $T_{\rm eff}$ ranging from 5054 K
    \citep{Fuhrmann2004} to 5200 K \citep{Luck2005}, $\log{g}$ ranging from 4.40
    \citep{Mishenina2013} to 4.72 \citep{Takeda2005}, and [Fe/H] ranging from
    $-0.18$ \citep{Ghezzi2010} to +0.06 \citep{Takeda2005}.
    We adopted the parameters of \cite{Fuhrmann2004} ($T_{\rm eff}$ = 5054 K,
    $\log{g}$ = 4.54, [Fe/H] = $-$0.09, and $\xi$ = 0.9 km\,s$^{-1}$).
    \cite{DiFolco2004} suggested an interferometric angular diameter of
    $\theta_{\rm LD}=2.148\pm0.029$ mas and an effective temperature of
    $T_{\rm eff}=5122$ K.
    H$_\alpha$ profile fitting with spectroscopic and interferometric methods
    result in quite consistent $\alpha_{\rm CMA}=2.2$.
    The best-fit $T_{\rm eff}=5045$ K also agrees with the photometric
    $T_{\rm eff}=5074$ K from the $b-y$ and $c_1$ \citep{Hauck1998}.

    HD\,39587 ($\chi^1$\,Ori, 54\,Ori, HR\,2047) is a variable star of RS CVn
    type, orbiting by a faint ($V_{\rm A}$ = 4.4, $V_{\rm B} = 8.8$) companion
    with a mass of $0.15\pm0.01\ M_\odot$ \citep{Irwin1992}.
    \cite{Boyajian2012a} measured its angular diameter $\theta_{\rm LD}=1.051\pm
    0.009$ mas with a precision of 0.9\%.
    This yield a radius of $0.979\pm0.009\ R_\odot$ with a Hipparcos parallax of
    $115.43\pm0.27$ mas \citep{vanLeeuwen2007}.
    \cite{Boyajian2012a} also gave its effective temperature as $T_{\rm eff} =
    5961$ K, which generally agrees with the spectroscopic temperature reported
    by \cite{Fuhrmann2004}.
    We found the best-fit $\alpha_{\rm CMA}=1.5$ for spectroscopic parameters and
    $\alpha_{\rm CMA}=1.9$ for interferometric parameters.

    HD\,122563 (HR\,5270) is the most metal-poor halo star ([Fe/H]$\sim-2.6$)
    that can be seen by the naked eye ($V=6.2$) in the night sky.
    It is the only giant in our program stars and therefore provides unique
    constraints.
    \cite{Mashonkina2011} presented a detailed analysis of the NLTE formation of
    iron lines for this star, and we used the same parameters ($T_{\rm eff} =
    4600$ K, $\log{g}=1.60$, [Fe/H] = $-$2.60, and $\xi$ = 1.95 km\,s$^{-1}$).
    Recent interferometric measurements with CHARA and PTI \citep{Creevey2012}
    gave its angular diameter as $\theta_{\rm LD}=0.948\pm0.012$ mas, and
    $T_{\rm eff} = 4598\pm41$ K, which is consistent with its spectroscopic
    temperature.
    We found the best-fit $\alpha_{\rm CMA}=1.0$ with H$_\alpha$ profile, while
    the fitting of H$_\beta$ and H$_\gamma$ gave satisfactory agreement with
    $\alpha_{\rm CMA}=0.5$.
    The $\alpha_{\rm CMA}$ is lower than that of the dwarfs, implying that the
    stellar convective energy transport become less efficient after stars
    evolved off the main sequence.

    HD\,217014 (51\,Peg, HR 8729) is a nearby ($\pi=64.07\pm0.38$ mass;
    \citealt{vanLeeuwen2007}) G5\,V star with the first extrasolar planet ever
    discovered orbiting a Sun-like host \citep{Mayor1995}.
    Spectroscopic analysis gave its effective temperatures as ranging from 5710
    K \citep{Maldonado2012} to 5832 K \citep{Ramirez2009}.
    The narrow-band photometric color indices $b-y=0.416$ and $c_1 = 0.371$
    \citep{Hauck1998} indicate a rather low $T_{\rm eff}=5661$ K
    \citep{Alonso1996}.
    The parameters adopted here ($T_{\rm eff}=5793$ K, $\log{g}=4.33$, [Fe/H] =
    +0.20, and $\xi$ = 0.95 km\,s$^{-1}$, \citealt{Fuhrmann1998}) agree for the
    effective temperature with the interferometric results of $T_{\rm eff}=5804$
    K and $\theta_{\rm LD}=0.748\pm0.027$ mas \citep{Baines2008}.
    We found a best-fit of $\alpha_{\rm CMA}=2.0$ with both parameters giving
    consistent results.

    HD\,45282 is a G\,0 metal-poor star ([Fe/H] $=-1.5$) at low Galactic latitude
    ($b=-3.9^\circ$).
    Previous spectroscopic analyses gave its $T_{\rm eff}$ as ranging from 5150 K
    \citep{Fulbright2000} to 5344 K \citep{Gratton2000}.
    Considering its high reddening with $E(B-V)\simeq0.82$ at this direction
    \citep{Schlegel1998}, we used the IRFM $T_{\rm eff}$ from
    \citet{Casagrande2010}, which is thought to be less affected by interstellar
    medium.
    The Balmer profile $\chi^2$ fitting gives $\alpha_{\rm CMA}=0.1$ for
    H$_\alpha$ and $\alpha_{\rm CMA}=0.5$ for H$_\beta$ with the best-fit
    $T_{\rm eff}=5310$ K.
    The error of 0.05 on $E(B-V)$ causes an uncertainty of $\sim300$ K on
    $T_{\rm eff}$, therefore the photometric temperature is not reliable.

    For the Sun we used the widely adopted parameters of $T_{\rm eff}=5777$ K,
    $\log{g}=4.44$, [Fe/H] = 0.0, and $\xi=0.90$ km\,s$^{-1}$ and for Vega
    $T_{\rm eff}=9550$ K, $\log{g}=3.95$, [Fe/H] = $-0.50$, and $\xi=2.00$
    km\,s$^{-1}$ \citep{Castelli1994}.
    The solar $\alpha_{\rm CMA}$ from the H$_\alpha$ Balmer line is 2.0, while a
    determination based on the solar flux suggests a value of 1.5--1.75.
    Although this is consistent within the error bars, it has to be noted that,
    for both the Sun and Procyon, SED fitting leads to a notably lower value of
    $\alpha_{\rm CMA}$ than Balmer-line fitting.

    \subsection{$\alpha_{\rm CMA}$}
    \begin{figure}[ht]
        \begin{center}
            \parbox{1\linewidth}{\includegraphics[width=9cm]{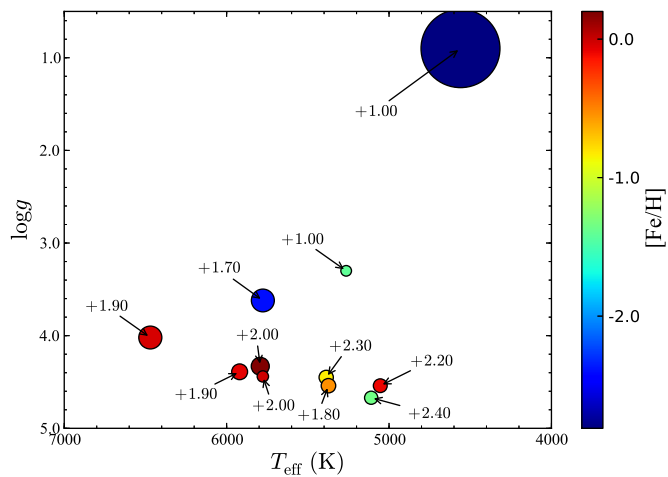}\\
            \centering}
            \hspace{1\linewidth}
            \hfill
            \\[0ex]
        \end{center}
        \caption{Determined $\alpha_{\rm CMA}$ for the program stars.
            An error of $\alpha_{\rm CMA}$ of $\approx\pm 0.4\dots 0.7$ maybe
            assumed for all stars except HD140283 ($T_{\rm eff} =5780$ K,
            $\log{g}=3.70$  where the error reaches $\approx\pm 1.1$.}
        \label{fig-alpha_result}
    \end{figure}

    Figure\,\ref{fig-alpha_result} shows our final results for the efficiency
    parameter $\alpha_{\rm CMA}$ as determined from Balmer-line fitting.
    Except for evolved stars, $\alpha_{\rm CMA}$ is consistent with a value
    of about 2.0 for all objects within the given insecurities.
    There is no obvious correlation with metallicity and a weak trend of
    increasing $\alpha$ for the coolest main-sequence objects in the investigated
    range of F- and G-type stars.
    This confirms the behavior and numerical values predicted by \cite{Magic2014}
    for this part of the HRD.

    For the evolved stars of our sample $\alpha_{\rm CMA}$ is of about 1.0.
    Although \cite{Magic2014} predicted the same trend of lower convective
    efficiency for objects with lower $\log{g}$, the magnitude of the change is
    much greater in our sample than that suggested by the 3D simulation of
    \cite{Magic2014}.
    Our sample mainly consists of main-sequence stars and only contains three
    evolved objects with well-determined parameters.
    Nevertheless, the stars clearly show a behavior different from that suggested
    by \cite{Magic2014}.
    The three objects span a wide range of temperatures, therefore the
    evolutionary state seems to be the main discriminator that separates them
    from the rest of the sample.

\section{Discussion and implication}

\begin{figure}[ht]
        \begin{center}
            \parbox{1\linewidth}{\includegraphics[width=9cm]{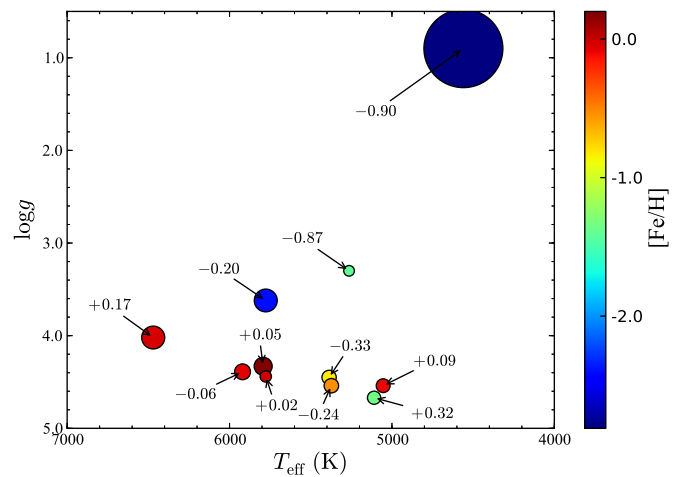}\\
            \centering}
            \hspace{1\linewidth}
            \hfill
            \\[0ex]
        \end{center}
        \caption{Difference between our determined $\alpha_{\rm CMA}$ and the
            one predicted by \cite{Magic2014}.}
        \label{fig-alpha_diff}
    \end{figure}

    Figure\,\ref{fig-alpha_diff} shows the difference between the suggested
    $\alpha_{\rm CMA}$ and that from \cite{Magic2014}, who adopted a calibration
    using the entropy jump (Fig.\,3 in their work).
    As described above, the differences are small along the main sequence and
    increase dramatically later in the evolutionary sequence, which suggests that
    there is either a stronger variation of $\alpha_{\rm CMA}$ as predicted by
    the STAGGER grid, or that our modeling of the Balmer lines is incorrect.
    The latter is excluded because we generally determined a temperature
    that agreed well with astrometric and IRFM data.

    Based on the similar method of FAG93, our analysis indicates that if the
    theory of \citet{Barklem2000} for Balmer-line broadening is adopted instead
    of that of \citet{Vidal1970, Vidal1973}, while the old ODF version is
    replaced by the MAFAGS-OS version, it is very unlikely to derive a low value
    of $\alpha_{\rm CMA}=0.82$.
    For F- and G-type main-sequence stars, $\alpha_{\rm CMA}\approx 2.0$ is
    derived, and the value decreases toward 1.0 for more evolved stars.

    A variation of the convective efficiency naturally influences the
    temperatures determined by the Balmer-line method.
    It furthermore changes stellar SEDs and in turn filter fluxes and color
    determinations.
    All stellar lines that have contribution functions down to the convection
    scone (very strong lines) are subject to potential changes as well.

    \vspace{2cm}
    

\begin{acknowledgements}
    We are grateful to K.Fuhrmann for providing some spectral observations.
    X.S.W. thanks R.Wittenmyer and H.N. Li for their useful comments and
    suggestions.
    X.S.W., L.W., and G.Z. are supported by the National Natural Science
    Foundation of China under grant No. 11390371 and 11233004.
    S.A. and L.M. are supported by the Russian Foundation for Basic Research
    (grants 14-02-31780 and 14-02-91153).
    F.G. would like to thank T.\,Gehren for his support, knowledgable comments,
    and so many helpful discussions.
\end{acknowledgements}

\bibliography{calibration}

\end{document}